\documentclass[useAMS,usenatbib]{mn2e}

\usepackage{graphicx}
\usepackage{url}
\usepackage{bm}
\usepackage{float}
\usepackage{placeins}
\usepackage{setspace}
\usepackage[fleqn]{amsmath}
\usepackage{longtable}
\usepackage{aas_macros}
\usepackage{amssymb}
\newcommand{\herschel}{{\it Herschel}}

\title{{\sc SEDeblend}: A new method for deblending spectral energy distributions in confused imaging}

\author[T. MacKenzie et al.]{Todd P. MacKenzie$^{1}$,
Douglas Scott$^{1}$,
Mark Swinbank$^{2}$
\\
$^{1}$Department of Physics \& Astronomy, University of British Columbia, 6224 Agricultural Road, Vancouver, BC V6T 1Z1, Canada \\
$^{2}$Institute for Computational Cosmology, Durham University, South Road, Durham, DH1 3LE, UK\\
}

\begin{document}
\maketitle

\begin{abstract}

For high-redshift submillimetre or millimetre sources detected with single dish telescopes, interferometric follow-up has shown that many are multiple submm galaxies blended together.  Confusion-limited {\it Herschel} observations of such targets are also available, and these sample the peak of their spectral energy distribution in the far-infrared.  Many methods for analysing these data have been adopted, but most follow the traditional approach of extracting fluxes before model spectral energy distributions are fit, which has the potential to erase important information on degeneracies among fitting parameters and glosses over the intricacies of confusion noise.  Here, we adapt the forward-modelling method that we originally developed to disentangle a high-redshift strongly-lensed galaxy group, in order to tackle this problem in a more statistically rigorous way, by combining source deblending and SED fitting into the same procedure.  We call this method ``SEDeblend.''  As an application, we derive constraints on far-infrared luminosities and dust temperatures for sources within the ALMA follow-up of the LABOCA Extended Chandra Deep Field South Submillimetre Survey.  We find an average dust temperature for an 870\,$\mu$m-selected sample of (33.9$\pm$2.4)\,K for the full survey.  When selection effects of the sample are considered, we find no evidence that the average dust temperature evolves with redshift.

\end{abstract}


\section{Introduction}

With the advent of single-dish submm observatories such as those using SCUBA-2 \citep{scuba2} on the JCMT, BLAST \citep{pascale2008}, and {\it Herschel} \citep{pilbratt2010}, we now have a window into the distant star-forming Universe \citep[e.g.][]{smail1997,barger1999,eales1999,scott2002,cowie2002,borys2003,coppin2006,patanchon2009,eales2010,elbaz2011,hermespaper,geach2013}.  However, due to the resolution of these observatories, instrumental noise is not the limiting factor when determining the uncertainty in flux density of individual sources that are observed for a sufficiently long period of time.  Instead, we are limited by confusion noise, caused by the high density of sources relative to the resolution of the imaging.  Higher resolution imaging can help with extracting the desired information from these confused images, but the current methods of combining such data are lacking in statistical rigour.  

A common exercise for learning about the physical properties of galaxies in these wavebands is to characterise the spectral energy distribution (SED) of a source.  When the source is much brighter than the confusion limit, this task is rather straightforward.  However, if the source is near or below the confusion limit for any particular waveband, then determining the SED becomes problematic.  This has been done with varying degrees of success using ``de-blending'' techniques \citep[e.g.][]{1deblend2005,2deblend2010,3deblend2011,swinbank}, often using positional priors from other higher-resolution observations to first extract fluxes, then subsequently fit SED models.  This two-step process does not usually fully account for the statistical properties of confusion noise (both spatial and between wavebands) and often neglects useful information regarding degeneracies among SED model fits with nearby sources, and thus the attribution of uncertainties to fit parameters becomes problematic.  We present here a method of combining high-resolution imaging with confused imaging, which simultaneously fits SEDs and separates sources, thus deblending SEDs instead of flux densities.  To do this, we adapt the forward-modelling method of \cite{Mackenzie2014} (henceforth referred to as M14) and generalise it to the case of point source deblending of model SEDs.  This new method forward-models each source SED to recreate the image plane and uses a Monte Carlo Markov chain (MCMC) Metropolis-Hastings algorithm \citep{mcmc1,mcmc2} with Gibbs sampling \citep{gibbs} to determine the uncertainties of the model parameters.  We apply our method to the Atacama Large Millimeter/submillimeter Array (ALMA) Survey of Submillimetre Galaxies \citep[ALESS,][]{alma} in the Extended Chandra Deep Field South (ECDFS) to measure the far-IR properties of the LABOCA ECDFS Submillimetre Survey selected sources \citep[LESS,][]{less}.  This task has already been undertaken by \cite{swinbank}, allowing us to compare our results with those of a more traditional method. Along with 870\,$\mu$m ALMA data, this region of the sky has also been imaged with the {\it Herschel} Spectral and Photometric Imaging Receiver \citep[SPIRE,][]{spire} and Photoconductor Array Camera and Spectrometer \citep[PACS,][]{pacs}, thus making it the ideal arena to test the effectiveness of our method.  Throughout we employ a $\Lambda$CDM cosmology with $\Omega_\Lambda=0.692$, $\Omega_\mathrm{m}=0.308$, and $H_0=67.8\,\mathrm{km}\,\mathrm{s}^{-1}\,\mathrm{Mpc}^{-1}$ \citep{planckcosmology}.

\section{A framework for fitting SEDs to blended sources}
\label{model}

%When creating a model, one must confront it with the data at some middle ground so that the data can be fit.  Traditionally, the telescope collects photons from the sky, which we use to reconstruct an image.  Sources are then identified and fluxes extracted, to which our SED model is then fit.  However, for confused images, source identification is problematic and flux extraction with deblending techniques erases important information, such as degeneracies among neighbouring sources and the correlated confusion noise.  The method presented here forward models the source positions and SEDs in order to reconstruct the image plane in each waveband, and thus retains this important information.  Instead of our SED model being fit to extracted fluxes, our model is fit to the images themselves and requires all source SEDs within a field to be fit simultaneously.  This is made possible only when higher resolution source positions are available from an independent source.

\subsection{Model SED and image reconstruction}
\label{sedmodelling}

We adopt a modified blackbody SED with a power-law component for the shorter wavelengths, as in M14.  Because we are not dealing with multiple images here (i.e. not strongly lensed), the image planes are reconstructed as follows:

\begin{equation}
M_{b}(\boldsymbol{x})=\sum\limits_i \bar{S}_{b}(T_{\mathrm{d},i}, z_i, C_i) P_\nu(\boldsymbol{x}-\boldsymbol{r}_{i}) + B_b.
\end{equation}

\noindent
Here $M_{b}(\boldsymbol{x})$ is the reconstructed image for frequency channel $b$, $\boldsymbol{x}$ denotes the position within the image, $\bar{S}_{b}$ is the source flux density of source $i$ averaged over the channel $b$ transmission filter, $T_{\mathrm{d},i}$ is the dust temperature, $z_i$ is the redshift, $C_i$ is a normalisation factor of source $i$, $P_\nu(\boldsymbol{x}-\boldsymbol{r}_{i})$ is the response function (i.e.~the telescope beam), with $\boldsymbol{r}_{i}$ denoting the position of source $i$, and $B_b$ is the image background.  The beam response functions for the {\it Herschel} channels are approximated as Gaussians with FWHM values of 11.6, 18.1, 24.9 and 36.2 arcseconds at $160$, $250$, $350$ and $500$\,$\mu$m, respectively \citep{spire}.

\subsection{\textit{Herschel}-SPIRE sky residuals}
\label{sky residuals}

\begin{figure*}
\begin{center}
\includegraphics[width=14cm]{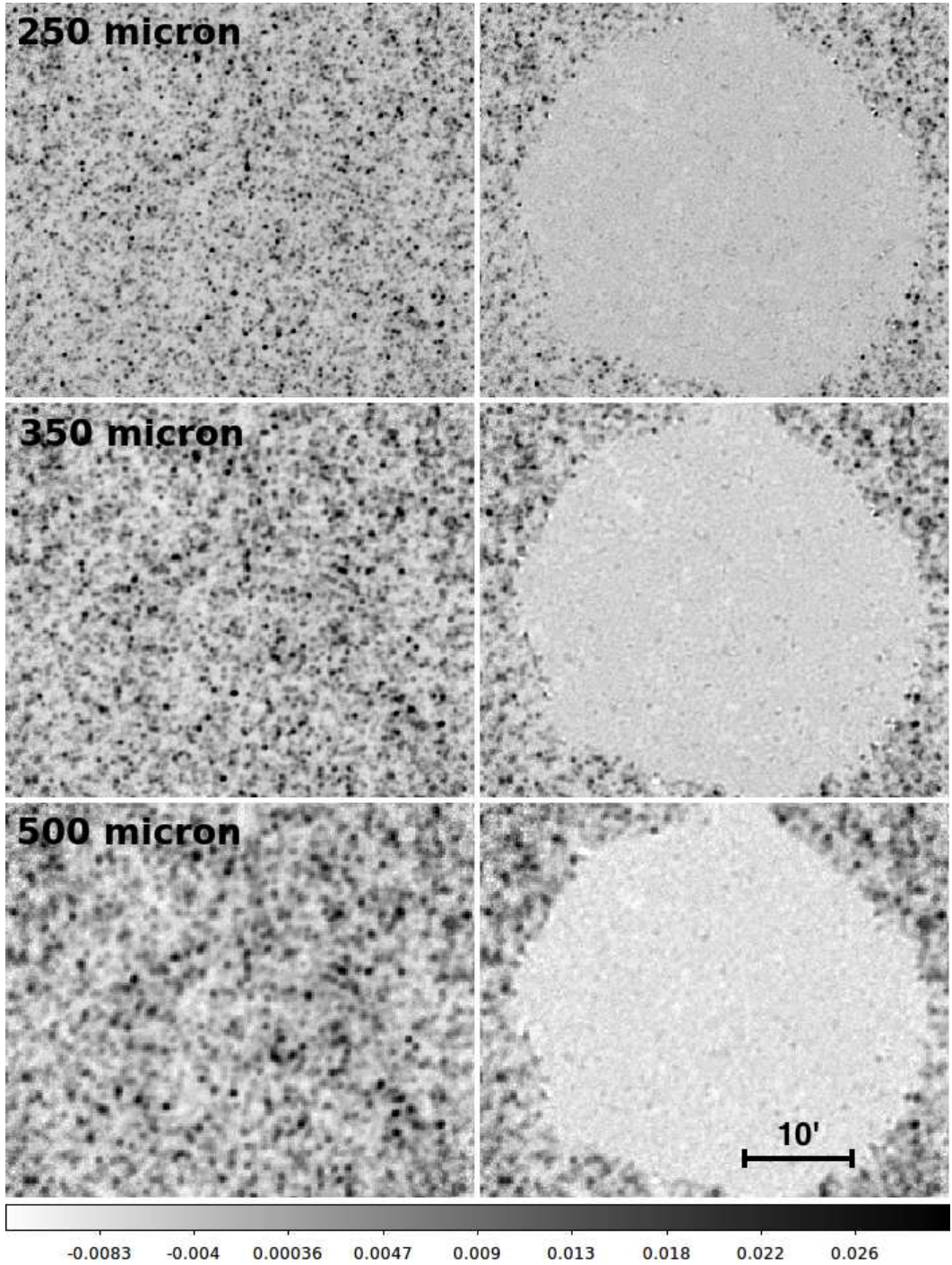}
\end{center}
\caption[{\it Left}: SPIRE ECDFS field.  {\it Right}:  SPIRE ECDFS field, after source subtraction.]{{\it Left}: SPIRE ECDFS field.  {\it Right}:  SPIRE ECDFS field, after source subtraction of 4024 mid-IR, radio and ALMA sources in the region of the sky where the 250\,$\mu$m instrumental noise is less than 1.2\,mJy.  The standard deviations of the residuals in this region after subtraction are 1.5, 1.6 and 1.4\,mJy at 250\,$\mu$m, 350\,$\mu$m and 500\,$\mu$m, respectively.  These residuals are still larger than the instrumental noise and are presumably dominated by sources too faint to be included in the catalogue of sources subtracted.  We will use these residuals to estimate the covariance of the sky when performing our full SED fitting.  The scale at the bottom of the image is in Jy.  The maps centres are located at roughly 3h32m30s $-$27$^\circ$47'00''.}
\label{subtraction}
\end{figure*}

In M14, additional deep cosmological field imaging was used to estimate the covariance of the sky in the likelihood calculation.  In this study, we are deblending the ALESS sources with the catalogue of nearby Multiband Imaging Photometer for Spitzer (MIPS) 24\,$\mu$m and Very Large Array (VLA) sources provided in \cite{swinbank}, henceforth referred to as the NMJS catalogue.  This catalogue accounts for the majority of the flux in the {\it Herschel}-SPIRE data (see below for details) and thus, using a cosmological field without subtracted sources to estimate the covariance for our likelihood calculations is not appropriate here.  Instead, we use the ECDFS SPIRE residuals, after subtracting our model SED, fit to every NMJS and ALESS source simultaneously.  

To achieve this, a maximum likelihood method is used to fit our model SEDs using a similar method to that described in Section~\ref{modelfitting}, weighting each pixel equally within the SPIRE data and ignoring any covariance between pixels.  Fig.~\ref{subtraction} shows the ECDFS field at the three \herschel{}-SPIRE wavelengths before and after the NMJS catalogue and ALESS sources are subtracted.  No PACS or ALMA data are used in this step.  We limit this process to the region where the 250\,$\mu$m instrumental noise is less than 1.2\,mJy, which includes 4024 sources from both the ALESS and NMJS catalogues.  Total flux densities from all NMJS and ALESS sources combined of 37.2, 28.6 and 16.2\,Jy are subtracted from the data at 250\,$\mu$m, 350\,$\mu$m and 500\,$\mu$m, respectively.  To test if we are over-subtracting flux from the maps, we stack the original maps on the positions of the catalogues, which produces total flux densities of 28.7$\pm$0.7, 23.4$\pm$0.6 and 14.5$\pm$0.5\,mJy at 250\,$\mu$m, 350\,$\mu$m and 500\,$\mu$m, respectively, with the errors estimated by bootstrapping.  One might conclude that we are over-subtracting, but stacking on the model sky (the image subtracted from the data to produce the residuals) produces total flux densities of 30.5, 24.0 and 14.0\,mJy at 250\,$\mu$m, 350\,$\mu$m and 500\,$\mu$m, respectively.  Both of these stacking results are significantly lower than the total flux densities of the subtracted sources, however we only expect stacked results to equal the total flux densities of the sources if they are Poisson distributed on the sky \citep{marsden2009}.  Because the stacking on the real and model sky give consistent results, we conclude that we are not significantly over-subtracting flux density from our maps.  Due to the effects of having a finite sized beam when creating the NMJS catalogue, where no two sources can occupy the same location on the sky, our resulting catalogue of sources is not Poisson distributed.  By generating two sets of simulated sky maps, one using the NMJS positions, and the second using random sky positions, where both use the same source flux densities, we are able to test this hypothesis.  The stacking on the simulated maps using the NMJS positions generated stacking results that had lower total flux densities than the sources used to generate the simulated maps, while stacking on the simulated maps using random source positions generated stacking results that equalled the total flux densities of the sources used to generated the simulated maps.

The standard deviations of the residuals are 1.5, 1.6 and 1.4\,mJy at 250\,$\mu$m, 350\,$\mu$m and 500\,$\mu$m, respectively; these values are significantly reduced from the confusion limits of 5.8, 6.3 and 6.8\,mJy, respectively \citep{confusion}.  Hence, a 24\,$\mu$m + 1.4\,GHz catalogue with signal-to-noise ratio $>$ 5 depths of 56\,$\mu$Jy and 41.5\,$\mu$Jy, respectively, accounts for approximately 80\% of the confusion noise in the maps.  These residuals are greater than the instrumental noise levels of 1.0, 1.1 and 1.2\,mJy in these regions; we are thus seeing the residual confusion noise of the sources that are not bright enough to be included in the ALESS and NMJS catalogue of sources we subtracted.  These residuals will be used in Section~\ref{modelfitting} to estimate the covariance of the sky.  This method allows us to greatly reduce the effects of confusion noise; instead, we are left with degeneracies in SED fitting parameters among the many nearby sources in our catalogues.  
%In theory, we could iterate on this source subtraction process, using our estimated covariance of the residuals to again subtract the same sources, using the full method described below.  Repeating this would converge to a more accurate view of the residual sky covariance, however this process would require at least three orders magnitude more computational time with marginal benefit.

\subsection{Model fitting}
\label{modelfitting}
As in M14, the model is fit to the data using an MCMC Metropolis-Hastings algorithm \citep{mcmc1,mcmc2} with Gibbs sampling \citep{gibbs}.  The $\log$ likelihood function for the {\it Herschel}-SPIRE data is 

\begin{equation}
-\log L_{\mathrm{SPIRE}}=X_{\mathrm{SPIRE}}+ \frac{1}{2}\bm{R}^\mathrm{T}\textsf{\textbf{C}}^{-1}\bm{R},
\end{equation}

\noindent
where $\bm{R}$ is a one-dimensional list of the residuals, and contains all three channels of SPIRE data ($\bm{R}=\{D_{250}(\boldsymbol{x}_{k})-M_{250}(\boldsymbol{x}_{k})/c_{250},D_{350}(\boldsymbol{x}_{k})-M_{350}(\boldsymbol{x}_{k})/c_{350},D_{500} (\boldsymbol{x}_{k})-M_{500}(\boldsymbol{x}_{k})/c_{500}\}$), $\textsf{\textbf{C}}^{-1}$ is the inverse covariance matrix for the residuals, $c_b$ are the calibration factors of each respective band, and $X_{\mathrm{SPIRE}}$ is a constant.  For each step in the MCMC chain, we are only interested in the differences between log-likelihoods, thus any constants can be ignored.

\begin{figure*}
\begin{center}
\includegraphics[width=12cm,trim = 10mm 0mm 0mm 0mm, clip, angle=-90]{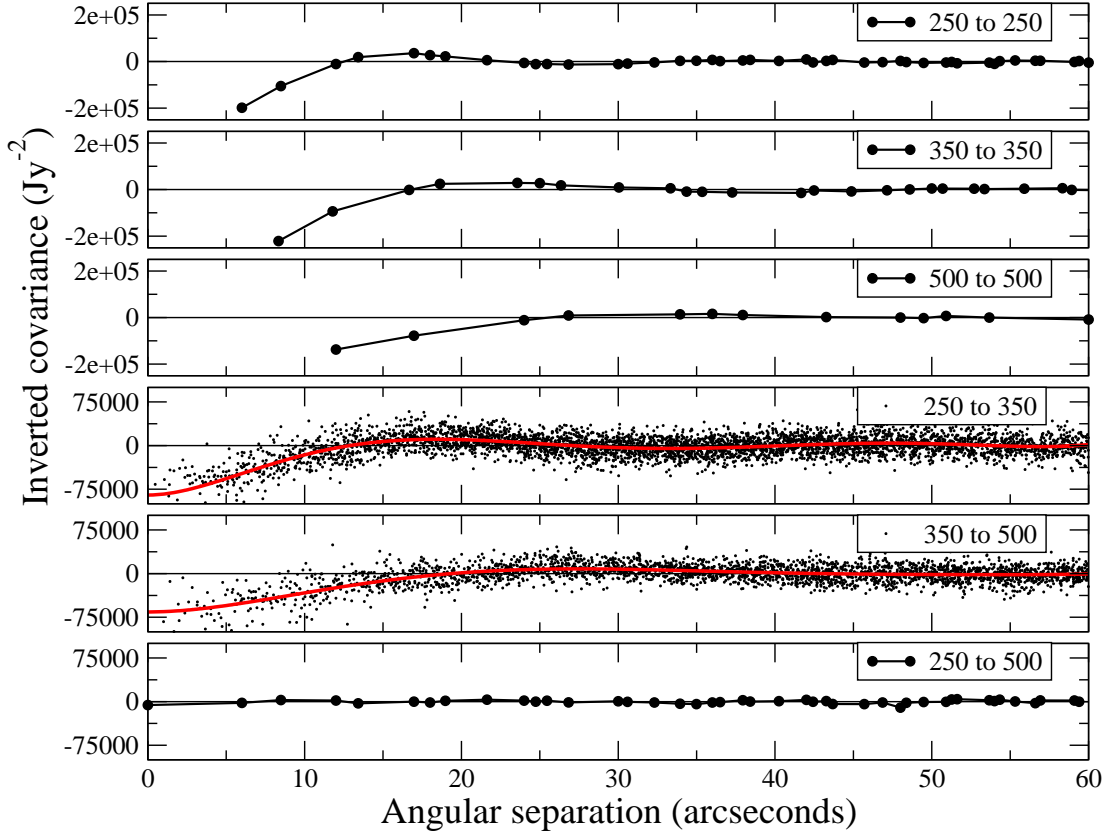}
\end{center}
\caption[Results of separating the inverted covariance matrix of the {\it Herschel}-SPIRE residuals by angular separation and by wavelength.]{Results of separating the inverted covariance matrix of the {\it Herschel} SPIRE residuals by angular separation and by wavelength.  The inverse auto-covariances for 250\,$\mu$m to 250\,$\mu$m, 350\,$\mu$m to 350\,$\mu$m, and 500\,$\mu$m to 500\,$\mu$m at an angular separation of zero are 1.02$\times10^6$\,Jy$^{-2}$, 1.07$\times10^6$\,Jy$^{-2}$, and 7.9$\times10^6$\,Jy$^{-2}$, respectively, and are not shown on the graphs above for clarity.  These inverted covariance lists are used in estimating the likelihood of our model fits.  From these lists/fits, we can see that covariances between neighbouring pixels contribute significantly to the likelihood estimate.}
\label{inverses}
\end{figure*}

In M14 the area of sky used was only a few arcminutes across, but the method described here must function on much larger areas, which is a problem, since the above calculation time scales with the square of the area used.  Fortunately, the covariance between pixels is only significant for nearby pixels, and so we do not need the whole matrix. We can estimate the covariance for an image of $10\times10$ pixels at each of the three SPIRE channels by selecting randomly chosen cutouts from the residuals described in Section~\ref{sky residuals}.  To perform this task, the covariance matrix is inverted and the result separated into six lists, corresponding to inverse covariances between pixels within the same waveband and between wavebands.  Where the angular distances between pixels forms a regular repeating pattern (due to the relative pixel sizes), we take the median inverse covariance value for each group of points with identical angular separations to obtain a better estimate of the inverse covariance.  For inverse covariances between 250\,$\mu$m and 350\,$\mu$m, and 350\,$\mu$m and 500\,$\mu$m, a high-order polynomial is fit to the data (the pixel sizes of 6, 8.3 and 12 arcseconds do not form simple repeating angular separation values between the wavelengths).  Fig.~\ref{inverses} shows the inverse covariance as a function of angular separation for the six lists.  If we limit the log-likelihood calculation to only pixels within a fixed radius of the sources of interest and between pixels within a fixed angular distance, the resulting likelihood calculation only scales with the area of sky used.

In theory, we could iterate on the process of making residual maps for use in estimating the residual sky covariance.  Where we treated each pixel with equal weight in Section~\ref{sky residuals}, we could instead use the estimated covariance from the previous iteration.  In practice however, the computational time of the likelihood calculation would become prohibitively large compared to the simple approach we implemented.  Fortunately, the residuals are likely dominated by sources too faint to be included in our NMJS catalogue, and not by a poorly weighted fit, and thus little would be gained by iterating on the residuals.  To test this assumption, we keep track of the total variance of the pixel values in the residual maps for each MCMC chain point while performing the full SED fitting (described below). We find an average minimum total variance of all the fields corresponding to 1.5, 1.6 and 1.4\,mJy standard deviations of the pixel values at 250, 350 and 500\,$\mu$m, respectively.

Flux calibration uncertainties, $c_b$, are taken into account during the fitting procedure by setting priors on $c_b$ for each band.  SPIRE waveband calibrations are correlated, with a covariance matrix

\begin{equation}
\textsf{\textbf{C}}_{\mathrm{cal}}=\left| \begin{array}{ccc}
  0.001825 & 0.0016 & 0.0016 \\
  0.0016 & 0.001825 & 0.0016 \\
  0.0016 & 0.0016 & 0.001825 
  \end{array}\right|, 
\end{equation}

\noindent
where the calibration is normalised to unity \citep{bendo}.  This corresponds to a 4\% correlated uncertainty between bands plus 1.5\% uncorrelated uncertainty between bands.  %The nominal pointing accuracy of the {\it Herschel} Space Telescope is 1.5 arcseconds, and thus we include this as a prior and marginalise over any possible pointing offset along with the image backgrounds.

The log-likelihood for the ALMA fluxes for a given band is given by

\begin{equation}
-\log L_{\mathrm{b}}=X_{\mathrm{b}}+ \sum\limits_i \frac{1}{2\sigma_{i,b}^2}(D_{i,b}-M_{i,b}/c_b)^2,
\end{equation}

\noindent
where $D_{i,b}$ is the measured flux density for source $i$, $\bar{M}_{i,b}$ is the model flux density for source $i$, $\sigma_{i,b}$ is the uncertainty in the measurement of $D_{i,b}$, $c_b$ is the calibration factor, $X_{\mathrm{b}}$ is a constant, and $b$ denotes the band of the measurement.  Unlike the \herschel{}-SPIRE bandpass filters, the ALMA bandpass filter is narrow and $\bar{M}_{i,b}$ is taken to be the flux density at the specified frequency.  The data used in this study are at 345\,GHz in ALMA Band 7, although we also consider the benefits of using additional 650\,GHz Band 9 data for constraining the far-IR properties of the ALESS sample in Section~\ref{degeneracies} (for future consideration).  Calibration uncertainties are 10\% and 20\% in Bands 7 and 9, respectively. %\citep[see][]{capabilities}. 

Because the 160\,$\mu$m PACS data are dominated by instrumental noise, the log-likelihood for these data is given by

\begin{multline}
-\log L_{\mathrm{PACS}}= \\ X_{\mathrm{160}}+ \sum\limits_{k} \frac{1}{2\sigma(\bm{x}_{k})^2}(D_{160}(\bm{x}_{k})-M_{160}(\bm{x}_{k})/c_{160})^2,
\end{multline}

\noindent
where $D_{160}(\bm{x}_{k})$ are the data, $M_{160}(\bm{x}_{k})$ is the sky model, $\sigma(\bm{x}_{k})$ is the instrumental error, $X_{\mathrm{160}}$ is a constant, $c_{160}$ is the calibration factor, and $\bm{x}_{k}$ is the position of pixel $k$ on the sky.  The 160\,$\mu$m PACS calibration uncertainty is 5\% \citep{muller2011pacs}.

\section{Testing with simulated sources}

While we do not require simulation of artificial sources in order to calibrate our method, we can use it as a tool to verify the accuracy of the uncertainties reported.  In particular, we can test how redshift, uncertainty in redshift, dust temperature and far-IR luminosity affect our ability to constrain these same properties.  We can also explore the effects of including nearby sources and the generated degeneracies among parameters.  In addition, we can quantitatively assess the benefits of adding further data, such as Band 9 ALMA measurements.

\subsection{Verifying our method}
\label{verify}

We verify our method by injecting simulated sources into the residual {\it Herschel} SPIRE images, described in Section~\ref{sky residuals}, along with simulated PACS data, and recording the resulting best-fit.  The best-fit distribution of the injected sources should match the expected uncertainties for such sources.  Simulated ALMA 870\,$\mu$m flux densities are given 0.5\,mJy Gaussian errors and the PACS 160\,$\mu$m data are simulated by generating a blank image with Gaussian random noise equal to the instrumental noise.  SPIRE calibration errors are randomly generated using the covariance matrix given in Section~\ref{modelfitting} and calibration errors for the ALMA and PACS data are also included.  This is, in effect, a Monte Carlo verification of our method and allows us to check the validity of our treatment of the {\it Herschel} SPIRE likelihood analysis.  We adopt a ``standard'' source with a redshift of 2, a dust temperature of 30\,K, and a far-IR luminosity of $10^{12}\,\mathrm{L}_\odot$, for the purpose of testing our method.  This equates to flux densities of 4.5, 6.4, 7.6, 5.6, and 1.8\,mJy at 160, 250, 350, 500, and 870\,$\mu$m, respectively, with a peak flux density of 7.7\,mJy at 323\,$\mu$m. We inject a total of 441 fake sources for each case we test below. Injecting each source one at a time allows us to test our constraining power for a single isolated source (although this is a rare occurrence due to the density of sources on the sky). To see the effect of source confusion, we can inject multiple simulated sources in close proximity.  Both of these cases are discussed below.  

\begin{figure}
\begin{center}
\includegraphics[width=8cm]{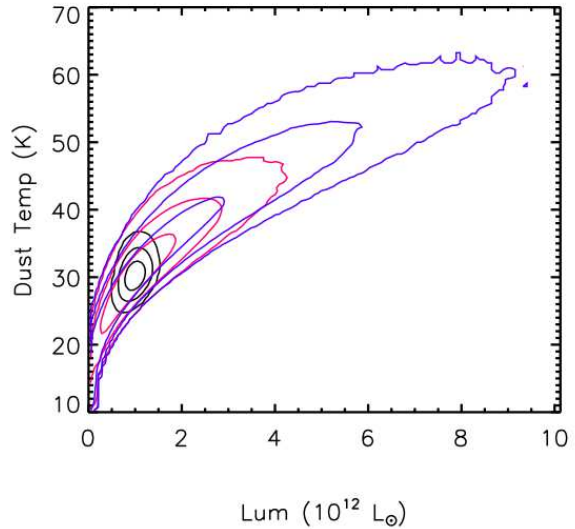}
\end{center}
\caption[Constraining power of our model for the case of varying redshift uncertainty.]{Constraining power of our model for the case of varying redshift uncertainty.  Here, 68, 95 and 99.7 percent credible regions are shown for redshift uncertainties of 0, $\pm$0.5 and $\pm$1, in black, red and blue contours, respectively.  A large uncertainty in redshift is one of the main limitations for constraining dust temperature, as well as far-IR luminosity.}
\label{redshifterror}
\end{figure}

Because dust temperature and redshift are entirely degenerate, one approach to take would be to constrain $T_\mathrm{d}/(1+z)$, instead of fixing the redshift and constraining dust temperature separately, as is done in most of the examples below.  However, because the ALESS sources have photometric redshift estimates from \cite{simpson}, it is beneficial to show constraints on dust temperatures separately.  The effect of an uncertainty in redshift is also explored below (see Fig.~\ref{redshifterror}).

\begin{figure*}
\begin{center}
\includegraphics[width=16cm]{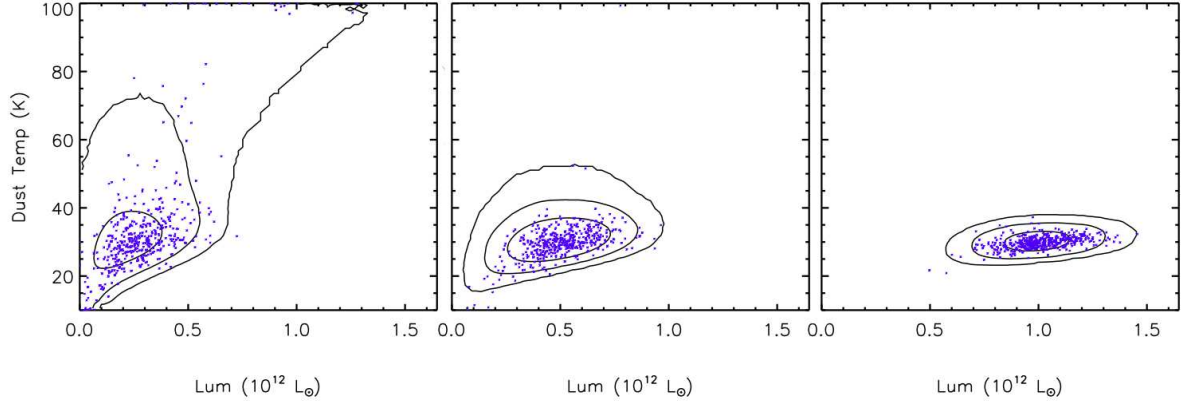}
\end{center}
\caption[Results comparing the expected uncertainty in fitting a source versus Monte Carlo simulated sources injected into the data.]{Results comparing the expected uncertainty in fitting a source given by our method (black contours), versus Monte Carlo simulated sources injected into the data.  The blue points are the Monte Carlo simulated sources used to verify our method.  In the right panel, we show our standard source with a redshift of 2, a dust temperature of 30\,K and a far-IR luminosity of $10^{12}\,\mathrm{L}_\odot$.  In the middle panel we show the same standard source with half the luminosity, and in the left panel, the standard source with a quarter of the original luminosity.  The black contours represent 68, 95 and 99.7 percent credible regions.  The Monte Carlo simulated sources trace out the expected uncertainties given by our method, thus we conclude that our likelihood analysis is validated.}
\label{testing}
\end{figure*}

Fig.~\ref{testing} shows the verification of our method for our standard source, as well as the cases where we decrease its luminosity by a factor of 2 and by a factor of 4.  Good agreement is found between our expected uncertainties and the Monte Carlo injected sources.  It is interesting to see the drastic change in temperature uncertainty as the luminosity of the standard source is reduced.  We could clearly provide constraints on sources to well below the confusion limit of the SPIRE data, if only we were dealing with isolated sources on the sky.  Of course this is just a tautology, since the sky is unfortunately a crowded place and our ability to constrain the properties of sources is largely limited by nearby sources that generate degeneracies in the fit parameters.

\begin{figure*}
\begin{center}
\includegraphics[width=12cm]{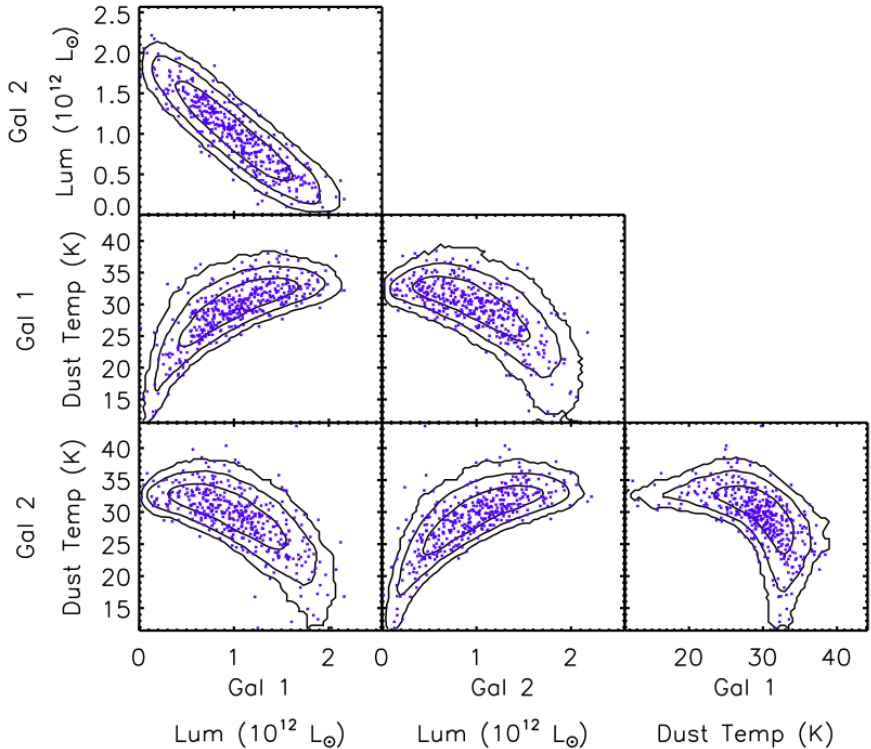}
\end{center}
\caption[Comparison of the expected uncertainties for two standard sources separated by 5 arcseconds given by our method (black contours), with the Monte Carlo simulated results (blue points).]{Comparing the expected uncertainties for two standard sources separated by 5 arcseconds with the Monte Carlo simulated results (blue points).  The black contours represent 68, 95 and 99.7 percent credible regions.  The Monte Carlo simulated sources trace out the expected uncertainties given by our method, providing an important validation of our likelihood analysis.  Also seen here are the degeneracies in fit parameters between the two neighbouring sources.  First, we can see that the far-IR luminosity of the two neighbouring sources are almost entirely degenerate, with the ALMA 870\,$\mu$m data providing most of the degeneracy-breaking power.  Second, we see that their dust temperatures are also anti-correlated, although the shape of the degeneracy here is more complicated.  Also seen here are degeneracies between the far-IR luminosity of one galaxy and the dust temperature of the second galaxy, and vice versa.  It is these degeneracies that would be erased if we employed a traditional two-step SED fitting of first deblending flux densities, with subsequent SED model fitting.}
\label{twosources}
\end{figure*}

Fig.~\ref{twosources} shows the verification of our method for the case of two standard sources separated by 5 arcseconds.  This example demonstrates a typical case of submm multiplicity as seen for many of the ALESS sources \citep{alma}.  Here, it is clear that the constraints on the properties of a source are limited by the degeneracies with its neighbour and not the residual unresolved far-IR background.  A linear anti-correlation between the two far-IR luminosities is expected, with the one-to-one degeneracy seen here the result of the two sources having the same far-IR luminosity and redshift.  The degeneracies seen between the other SED model parameters, typically ``banana-shaped,'' depend on the values of the parameters themselves.  It is these degeneracies that two-step SED fitting erases.  Again, our Monte Carlo simulated sources accurately reflect the expected uncertainties.

\begin{figure}
\begin{center}
\includegraphics[width=8cm]{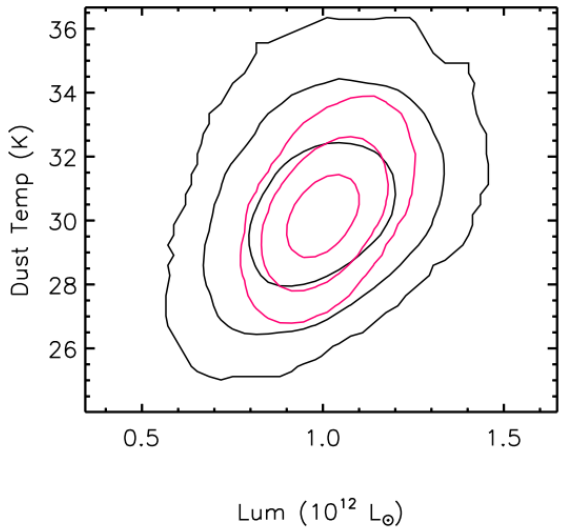}
\end{center}
\caption[Comparison of the expected uncertainties for a standard source using our method and an identical method with a naive approach.]{Comparing the expected uncertainties for a standard source using our method (black contours show 68, 95 and 99.7 percent credible regions) and an identical method with a naive approach of the {\it Herschel} SPIRE likelihood that considers only the instrumental noise in each pixel and ignores the covariance with neighbouring pixels (red contours).}
\label{nointerpixel}
\end{figure}

Fitting large numbers of Monte Carlo simulated sources is a computationally expensive exercise and thus we stop the verification of our method here.  We have shown that the constraints produced by our method accurately reflect the results of Monte Carlo simulations, and thus our treatment of the SPIRE likelihood analysis is validated.  For a standard source, Fig.~\ref{nointerpixel} shows the difference between our approach and an identical method where we only consider the instrumental noise of the SPIRE data and ignore the correlations between neighbouring pixels in angular separation as well as between wavelengths.  Without proper treatment of the SPIRE likelihoods, it is clear that we would be over-constraining the properties of sources within our model.

Assigning an accurate dust temperature uncertainty is an issue that has been neglected in much of the literature.  Constraints on dust temperature are affected by several factors, such as the redshift of the source, the width of the telescope bandpass filters, the wavelength coverage of the telescope filters, and the signal-to-noise ratio of the source within the images.  However, the dust temperature uncertainty naturally falls out of the method employed here, and thus we perform a few tests as examples.     

\begin{figure*}
\begin{center}
\includegraphics[width=12cm]{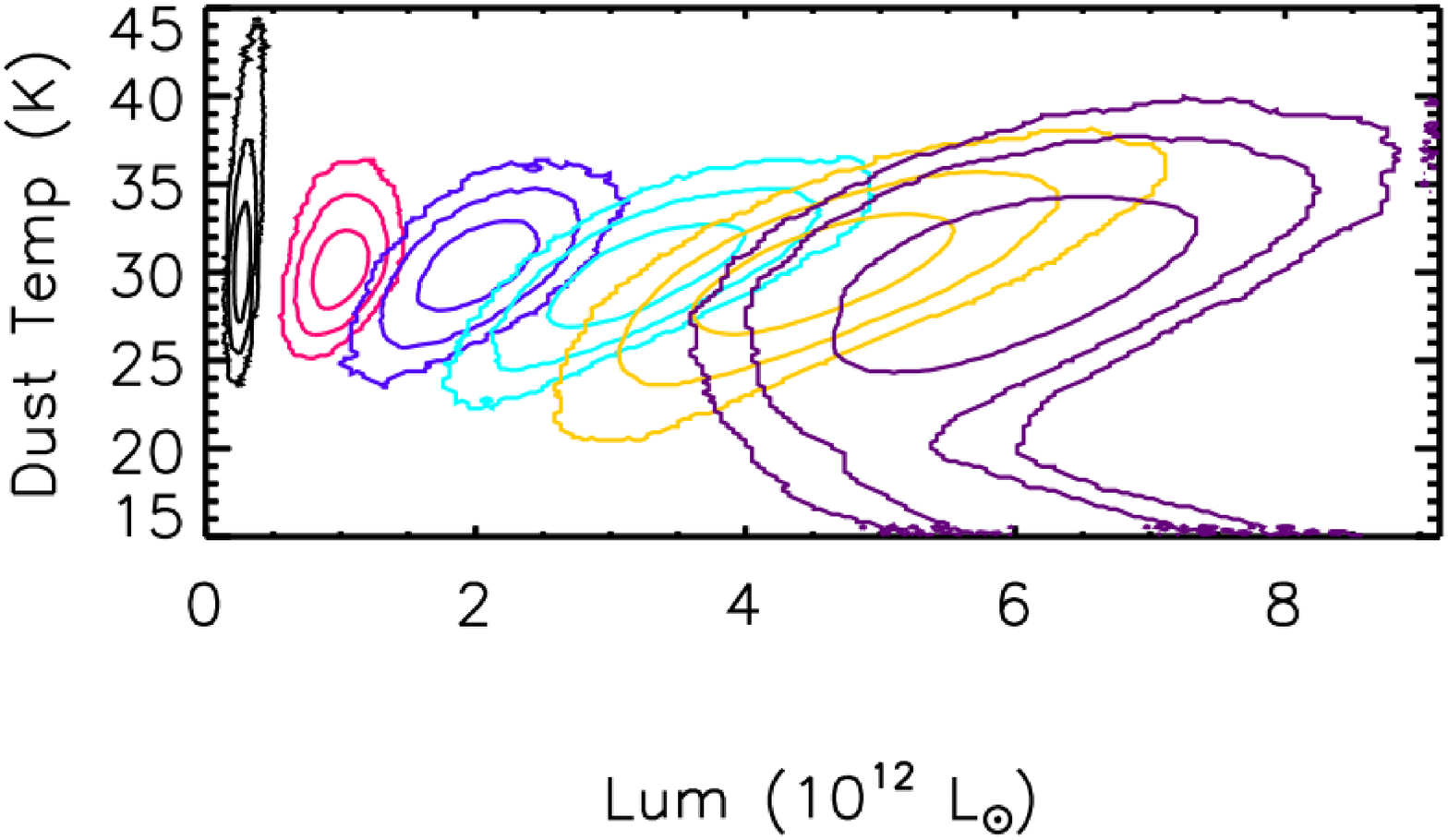}
\end{center}
\caption[Constraining power of our model as a function of redshift for our standard source while keeping peak flux density constant.]{Constraining power of our model as a function of redshift for our standard source while keeping peak flux density constant.  We show 68, 95 and 99.7 percent credible regions for redshifts of 1, 2, 3, 4, 5 and 6 in black, red, blue, green yellow and purple, respectively.}
\label{redshiftconstflux}
\end{figure*}

Fig.~\ref{redshiftconstflux} shows how our constraints change as we vary the redshift of our standard source while keeping the peak flux density constant and letting the far-IR luminosity change.  An interesting effect is seen at $z\gtrsim6$, where a colder fit to the dust temperature starts to increase the far-IR luminosity. This is because the peak of the SED shifts beyond the ALMA 870\,$\mu$m waveband.  A similar effect is seen at low redshifts, when the peak of the SED shifts to wavelengths shorter than 160\,$\mu$m and the upper dust temperature bound starts to rise.  For a dust temperature of 30\,K, these effects do not become significant unless the redshift is lower than about 1 or greater than about 6; thus the wavelength coverage of the available data is ideally suited for the sample of ALESS sources we are fitting in Section~\ref{alesssurvey}.

Up to this point, we have assumed that the redshift of our standard source was well constrained.  To investigate dropping this assumption, fig.~\ref{redshifterror} shows our model constraints for redshift uncertainties of 0, $\pm$0.5, and $\pm$1.  How well we can constrain dust temperature and far-IR luminosity, along with degeneracies among nearby sources, strongly depends on the uncertainty in source redshift.

\subsection{The addition of a second ALMA frequency}
\label{degeneracies}

\begin{figure*}
\begin{center}
\includegraphics[width=12cm]{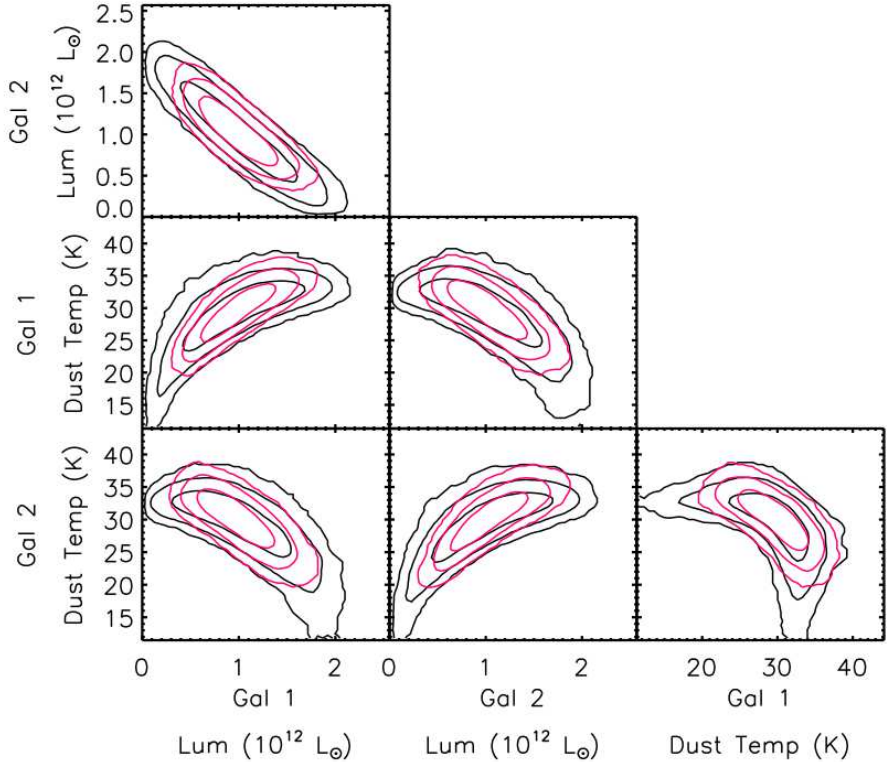}
\end{center}
\caption[Constraining power of our model for the case of two standard sources separated by 5 arcseconds.]{Constraining power of our model for the case of two standard sources separated by 5 arcseconds.  The black contours denote 68, 95 and 99.7 percent credible regions using 0.5\,mJy rms 870\,$\mu$m ALMA Band 7 observations, while the red contours are when 1\,mJy rms 460\,$\mu$m ALMA Band 9 observations are added along with 870\,$\mu$m ALMA Band 7 observations.}
\label{multiplicity}
\end{figure*}

\begin{figure*}
\begin{center}
\includegraphics[width=12cm]{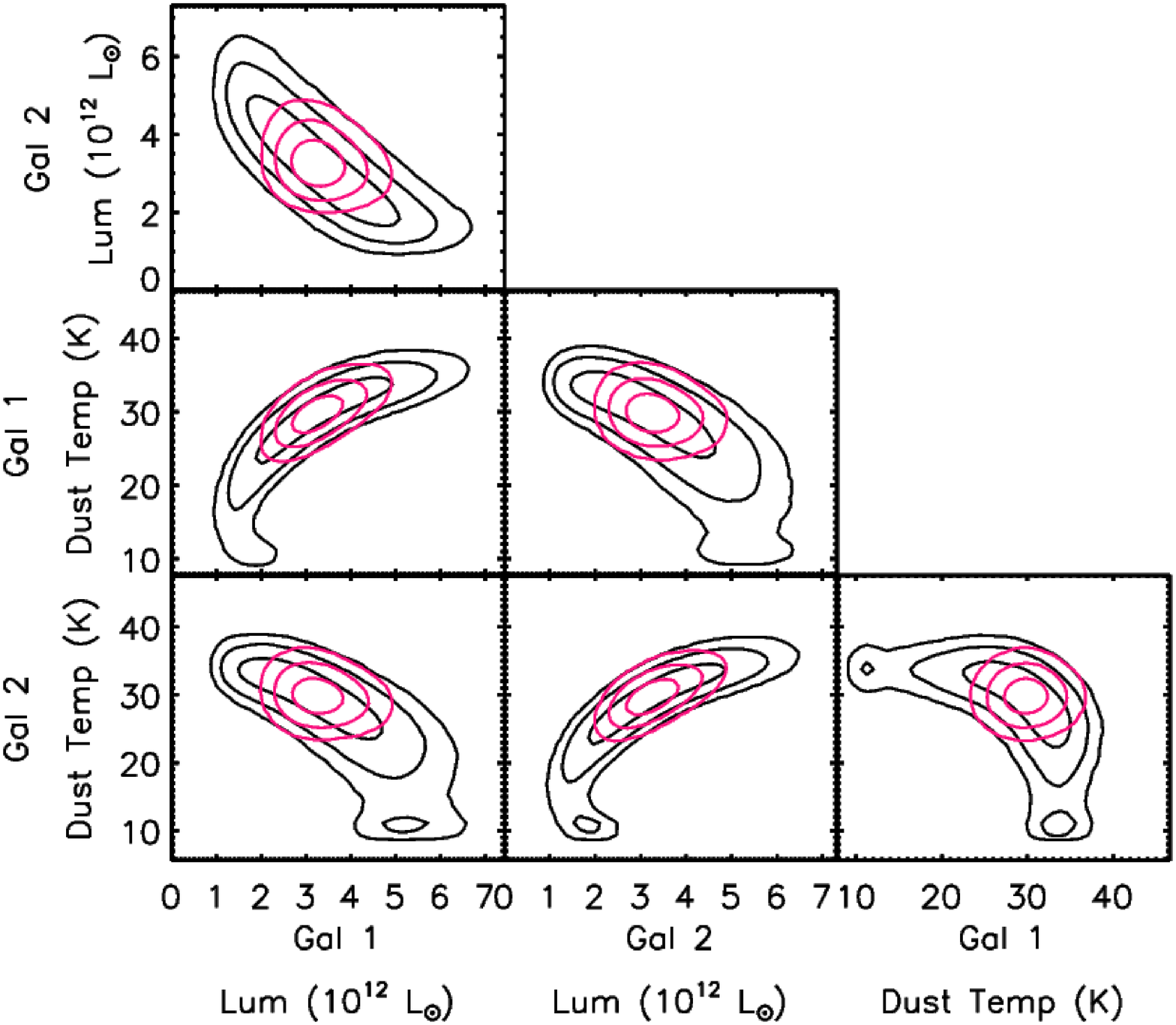}
\end{center}
\caption[Constraining power of our model for the case of two standard sources, moved to a redshift of 4, while keeping the same peak flux density, and a separation of 5 arcseconds.]{Constraining power of our model for the case of two standard sources, moved to a redshift of 4, while keeping the same peak flux density, and a separation of 5 arcseconds.  The black contours denote 68, 95 and 99.7 percent credible regions using 0.5\,mJy rms 870\,$\mu$m ALMA Band 7 observations, while the red contours are when 1\,mJy rms 460\,$\mu$m ALMA Band 9 observations are added along with 870\,$\mu$m ALMA Band 7 observations.}
\label{twosourcesz4}
\end{figure*}

ALMA follow-up observations of 870\,$\mu$m sources selected from ALESS \citep{less,alma} have shown that a significant fraction of single-dish detected sources are in fact comprised of multiple galaxies.  Since degeneracies with nearby sources are a dominating factor in determining our ability to constrain their far-IR properties (see Fig.~\ref{twosources}), such sources will have particularly poor constraints on their far-IR properties.  In Fig.~\ref{multiplicity} we explore the benefits of adding ALMA Band 9 observations at 460\,$\mu$m, with an rms of 1\,mJy, for the case of two standard sources separated by 5 arcseconds on the sky.  Since the peak of the SED for our standard source is at 323\,$\mu$m, which is shorter than both the ALMA wavelengths considered, only moderate improvement in constraining power is expected, and this is what is seen in the simulations.  Specifically, the lower bound on the temperature is improved, which in turn improves the constraint on far-IR luminosity.  Much greater improvements in constraining power are realised when the peak of the SED is straddled by the two ALMA wavelengths, as would be the case if our standard source were at a higher redshift or had a lower dust temperature.  Fig.~\ref{twosourcesz4} shows the improvement for the case of two standard sources separated by 5 arcseconds, where the standard sources are moved to a redshift of 4 and their peak flux densities remain unchanged.  In this case, degeneracies between the two sources are nearly eliminated when adding a second ALMA band.

\section{The properties of submm galaxies within the ALESS survey}
\label{alesssurvey}

When fitting our model to the data, we use the ALMA 870\,$\mu$m fluxes and positions from \cite{alma} and the photometric redshift estimates of \cite{simpson}, which were used by \cite{swinbank}.  The photometric redshift constraints are considered to be $\pm1\,\sigma$ Gaussian priors in our model.  As in \cite{swinbank}, we treat any source in the NMJS catalogue as a duplicate if it is within 1.5 arcseconds of an ALESS source or another NMJS source.  

We found that our data have almost no constraining power on the dust emissivity index, $\beta$, when it is allowed to range over 1.0$-$2.5, and thus we simply fix it to a nominal value of 1.5 so that we may easily compare the ALESS sample with the sample of \cite{symeonidis}.  We set a hard prior on the dust temperature such that it must be above 10\,K, since no colder galaxies have been found in any similar surveys \citep[e.g.][]{kingfish, herschelSEDpaper, symeonidis}, besides which the microwave background temperature sets a similar limit at a redshift $\sim$\,3.  This hard prior is useful for when the peak of the SED is shifted close to, or beyond, the ALMA 870\,$\mu$m wavelength, which occurs at high-redshifts when the source is cold (see Fig.\ref{redshiftconstflux}); thus this prior keeps the model from entering an unphysical region of parameter space.  We also use a hard prior to keep the dust temperature from going beyond 100\,K, since no source in the ALESS sample was found to be this hot by \cite{swinbank}.

We provide  the median values of our MCMC chains and report 68 percent credible intervals throughout.  Far-IR luminosities are calculated by integrating the model SED from 8 to 1000\,$\mu$m.  When either the dust temperature or far-IR luminosity  lower credible interval are consistent with either zero far-IR luminosity or 10\,K for dust temperature, we report the 84 percent credible interval as an upper limit.  Note that because of our prior on dust temperature, upper limits for dust temperature are somewhat subjective in that the upper limit would move if we changed the dust temperature prior.  While we may only have upper limits in one of these parameters, this does not necessarily translate into an upper limit on the other.  In fact, in only one case do we have an upper limit on both far-IR luminosity and dust temperature.   The resulting far-IR luminosity and dust temperature constraints are given in Table~\ref{propertytable}.  Note that we do not report any constraints for ALESS083.4, since the redshift of the source puts the peak of the SED at shorter wavelengths than the available data and thus no constraint on temperature is possible.

\subsection{Comparison with \protect\cite{swinbank}}

The benefit of applying our method to this sample of ALESS sources is that we can compare our results with those of \cite{swinbank}, who employed a simpler competing method of deblending and SED fitting.  To facilitate the comparison, we have used much of the same data, although there are also key differences that make a detailed comparison less straightforward.  To facilitate the comparison, we have used the same ALESS catalogue of positions and flux densities \citep{alma}, the same NMJS catalogue, the same {\it Herschel}-SPIRE and PACS 160\,$\mu$m data, and the same redshift estimates \citep{simpson}.  Aside from the method used to deblend the {\it Herschel} data, important differences in \cite{swinbank} includes the use of an SED library and the inclusion of both shorter and longer wavelength data when fitting SEDs.

\begin{figure*}
\begin{center}
\includegraphics[width=10cm,trim = 25mm 5mm 15mm 10mm, clip, angle=-90]{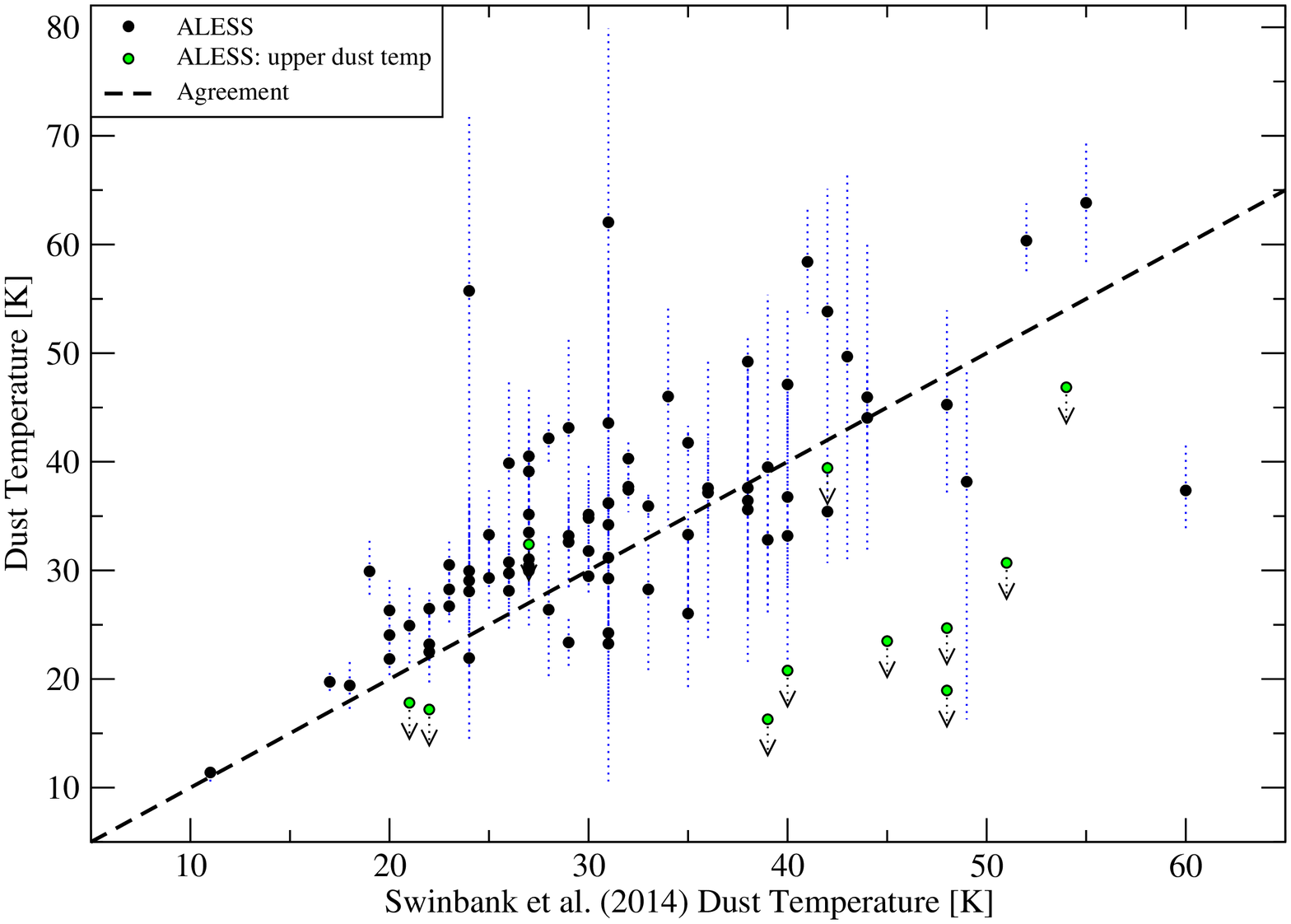}
\includegraphics[width=10cm,trim = 25mm 5mm 15mm 5mm, clip, angle=-90]{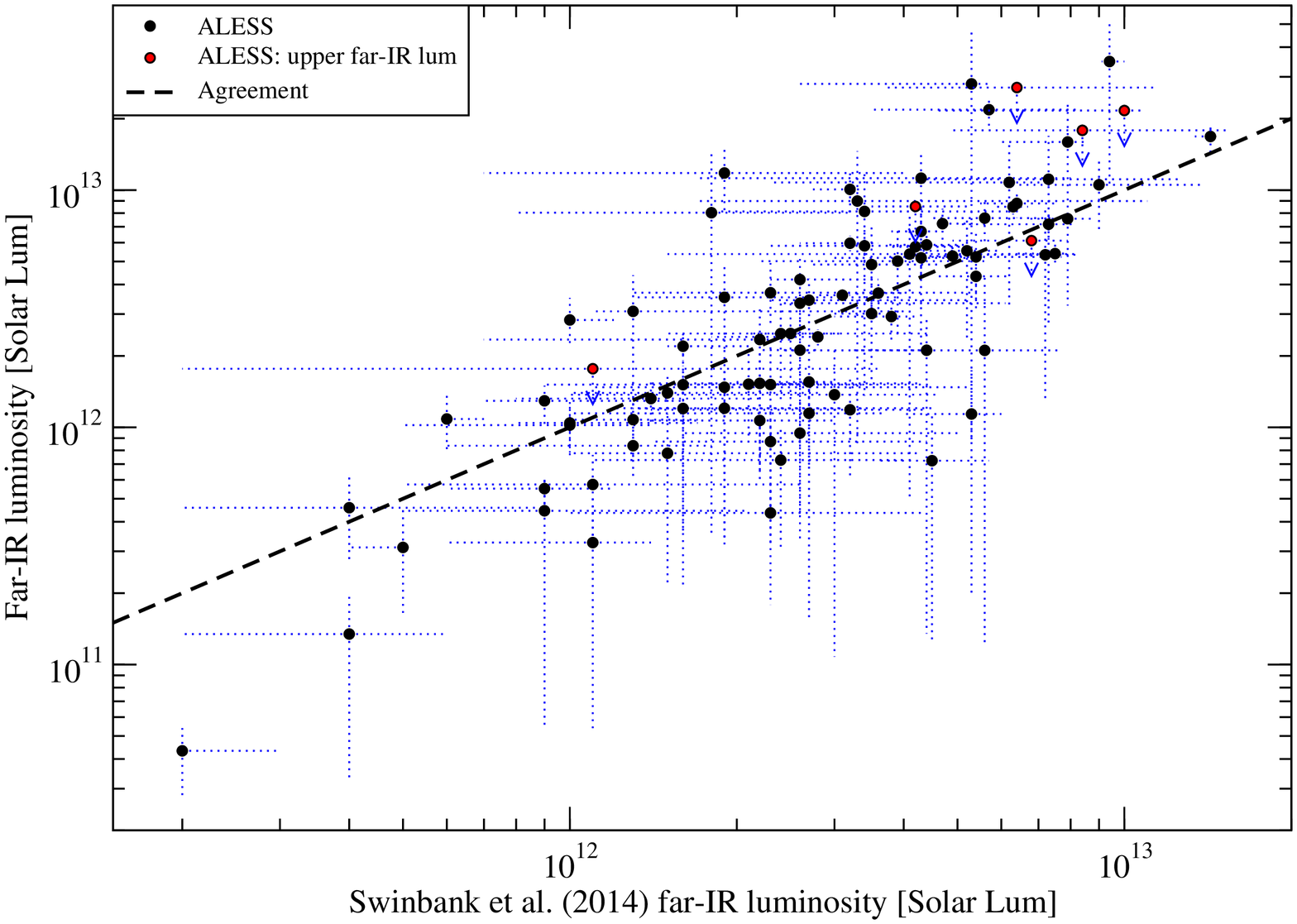}
\end{center}
\caption{Comparison of our results with those found by \protect\cite{swinbank}.  {\it Top}: comparison of dust temperatures between the two methods.  {\it Bottom}: comparison of far-IR luminosities.  Black points are ALESS sources with constraints on both the far-IR luminosity and dust temperature.  Red points are ALESS sources with 84 percent upper limits on far-IR luminosity.  Green points are ALESS sources with 84 percent upper limits on dust temperature.  The dashed black line shows the expected relation if the two methods were in agreement.  While our results show a clear correlation with those found by \protect\cite{swinbank}, there is disagreement for many of the individual ALESS sources.  One prominent feature appears to be roughly a 4\,K offset in temperature between the two methods.  This discrepancy is easily explained by the choice of dust emissivity index.  We have used a fixed dust emissivity of $\beta=1.5$, while \protect\cite{swinbank} used a value of $\beta$ that varied, with an average of 1.8.  This difference in dust emissivity index accounts for most of the apparent dust temperature shift.}
\label{comparison}
\end{figure*}

Fig.~\ref{comparison} compares the results of our two methods to assess their level of agreement.  The black dashed line in both plots shows the locus representing complete agreement, while the \cite{swinbank} dust temperatures used in the comparison are those that were derived from fitting a modified blackbody to the {\it Herschel} photometry.  We use a fixed dust emissivity index of 1.5, primarily so that we may also compare our results with those of \cite{symeonidis}.  An apparent systematic shift towards warmer dust temperatures is seen for our results, with an amplitude around 4\,K; however, comparing dust temperatures requires knowledge of the SED model used to fit the data and any priors on the dust emissivity index, $\beta$.  We found that using a dust emissivity index of $\sim1.9$ would eliminate this systematic shift, however \cite{swinbank} allowed the dust emissivity index to vary between 1.5 and 2.2 and found an average best fit value of 1.8, thus this dust temperature discrepancy is easily explained.  When we allowed the dust emissivity index to vary freely between 1 and 2.5, we found that the data had almost no constraining power on the value of $\beta$.  

%Another possible contribution in the dust temperature discrepancy lies within a deblending rule used by \cite{swinbank} and others \citep[e.g.][]{2deblend2010,3deblend2011}, where sources are included in the catalogue of sources to be deblended, only if they were found to have a flux density greater than 2\,$\sigma$ when deblended at the previous shorter wavelength.  For example, a nearby source is only included in the catalogue of sources to deblend at 350\,$\mu$m if it was found to have a significant flux density at 250\,$\mu$m (and only deblended at 500\,$\mu$m if it was found to have a significant flux density at 350\,$\mu$m).  This rule is used eliminate faint sources from the catalogue of sources to deblend when deblending longer wavelengths, where the resolution of the imaging is worse, and thus the number of sources per beam is higher.  The consequence of this rule is that cold or high redshift sources, which can have flux densities that increase with wavelength, are likely to be dropped from the list of sources to deblend.  Their flux densities at longer wavelengths could then be incorrectly attributed to neighbouring nearby sources, making them appear colder than they really are.

When comparing the far-IR luminosities, a clear correlation can be seen between the two methods, with a slight tendency for our new approach to fit higher far-IR luminosities for more luminous objects and lower far-IR luminosities for less luminous objects.  Again, the choice of specific SED model will affect results here, primarily the lack of a shorter wavelength hot component to our SED model, as well as the use of shorter and longer wavelength data used by \cite{swinbank}.  Such a comparison would require us to develop a more complicated SED model that would allow us to incorporate these other wavelengths.

Overall, we believe our method to be an improvement over what has been used in previous studies of submm galaxies and its effectiveness has been shown in Section~\ref{verify}.  In particular, it forgoes the need to deblend confused imaging prior to fitting SEDs. Our method fits SEDs and deblends the images simultaneously and can easily incorporate prior knowledge of the expected source SED shape.

\begin{figure*}
\begin{center}
\includegraphics[width=10cm,trim = 25mm 5mm 15mm 10mm, clip, angle=-90]{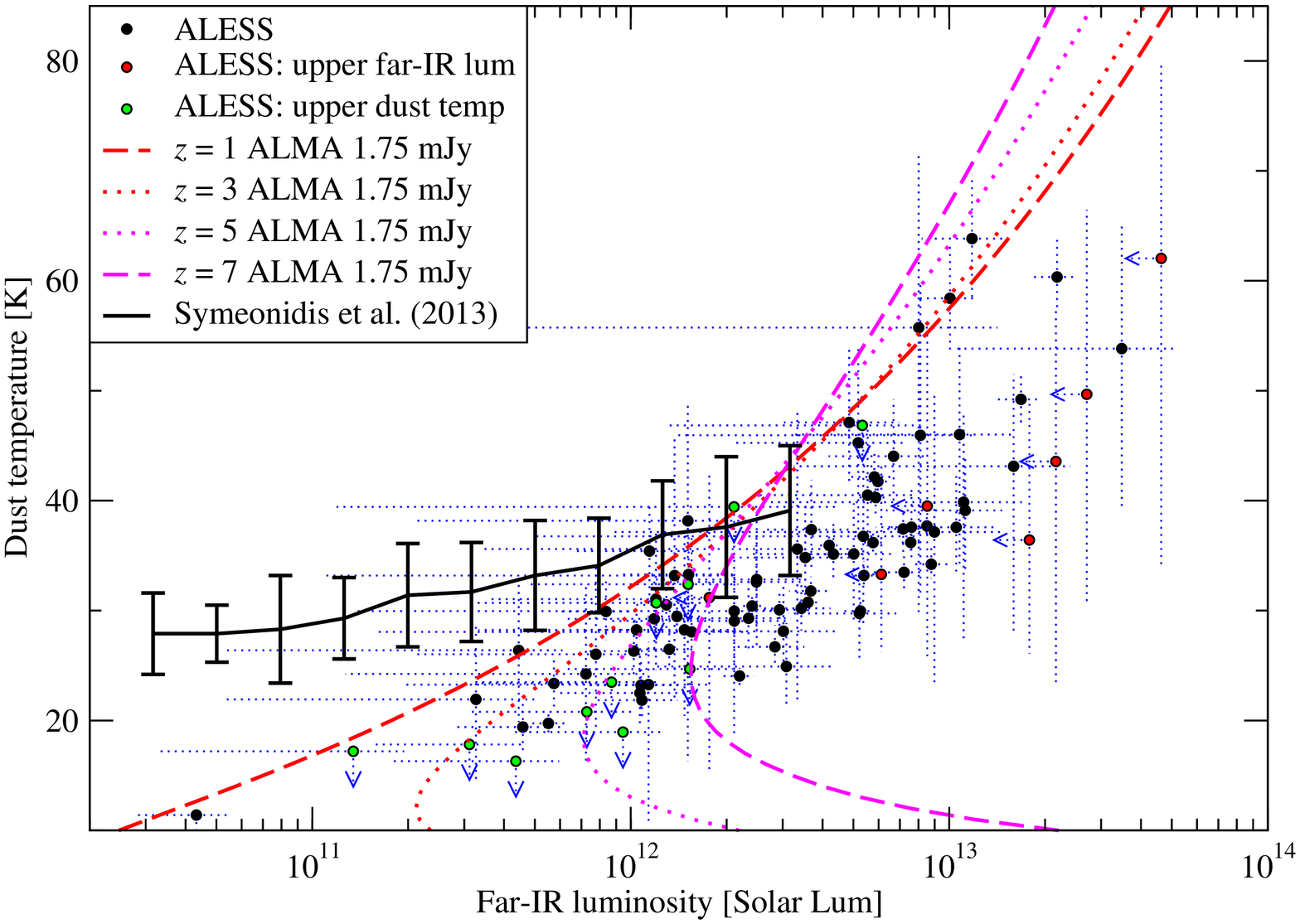}
\includegraphics[width=10cm,trim = 25mm 5mm 15mm 5mm, clip, angle=-90]{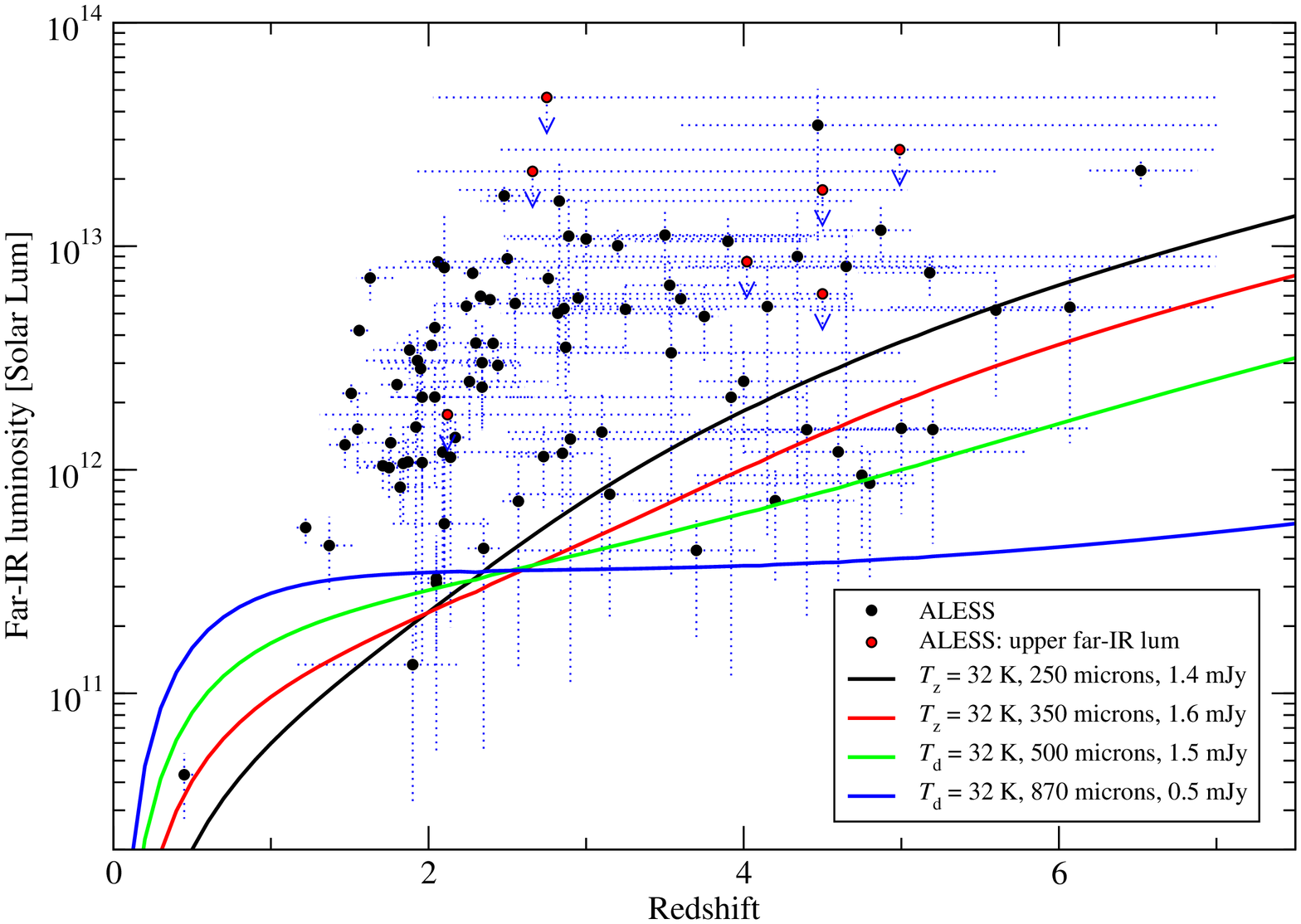}
\end{center}
\caption[{\it Top}:  dust temperature versus far-IR luminosity for the ALESS sample. {\it Bottom}:  far-IR luminosity versus redshift for the ALESS sample.]{{\it Top}:  dust temperature versus far-IR luminosity for the ALESS sample.  Black points are ALESS sources with constraints on both the far-IR luminosity and dust temperature.  Red points are ALESS sources with 1\,$\sigma$ upper limits on far-IR luminosity.  Green points are ALESS sources with 1\,$\sigma$ upper limits on dust temperature.  Dot and dashed lines are representative 3.5\,$\sigma$ detection limits of the ALMA data for redshifts between 1 and 7.  The solid black line is the far-IR luminosity to dust temperature relation found by \cite{symeonidis}.  It is clear from the detection limits that our sample is biased towards colder dust temperatures.  {\it Bottom}:  far-IR luminosity versus redshift for the ALESS sample.  The colour of the points denote the same objects as above.  Representative 1\,$\sigma$ detection limits are drawn for a $T_\mathrm{d}$\,=\,33\,K source at 250, 350, 500 and 870\,$\mu$m in black, red, green, and blue, respectively.  ALESS sources with upper limits on dust temperature can be found in the region between the ALMA and {\it Herschel} SPIRE detection limits, implying a detection by ALMA, but little or no flux seen by {\it Herschel} SPIRE.}
\label{results}
\end{figure*}

\subsection{Dust temperatures and selection effects for ALESS sources}

The top panel of Fig.~\ref{results} plots the dust temperature versus far-IR luminosity for the ALESS sample.  Many previous studies showed a correlation between dust temperature and far-IR luminosity, the $L$--$T_\mathrm{d}$ relation \citep[e.g.][]{chapman2005,magnelli2012,casey2012,symeonidis}, although some authors have noted that many of these studies are biased by selection effects \citep[e.g.][]{chapin2009,chapin2011,swinbank}.  Over-plotted on the top panel of Fig.~\ref{results}, using a solid black line, is the $L$--$T_\mathrm{d}$ relation as found by \cite{symeonidis}. The sample of sources used to find this relation were specifically chosen with the aim of minimising selection effects and are likely to be the most accurate representation of the low redshift $L$--$T_\mathrm{d}$ relation in the literature.  A major result of \cite{symeonidis} is that sources at $z<0.1$ are on average a few Kelvin warmer than those with redshifts ranging from 0.1 to 2.  For our study, we have specifically chosen a value of the dust emissivity index that allows us to compare our results directly to those of \cite{symeonidis}, to test if dust temperature evolves further at higher redshifts.  Upon first inspection, it would appear that the ALESS sources are indeed cooler; however, we must consider the selection effects of our sample.  In the top panel of Fig.~\ref{results}, the red and purple, dotted and dashed lines, denote representative ALMA 3.5\,$\sigma$ detection limits for redshifts of 1, 3, 5, and 7.  In the region where our two samples overlap, it is clear that these detection limits bias our sample to cooler temperatures.

%Many have claimed that high-redshift sources have a colder average dust temperature than those nearby ($z<0.1$) ****citations needed****.  For the ALESS sample of sources with far-IR luminosities between $10^{12}$ and $10^{13}\,L_\odot$, we find a bootstrapped dust temperature of 34.6$\pm$1.5\,K for sources with luminosities between $10^{12}$ and $10^{13}\,L_\odot$.  The top panel of Fig.~\ref{results} plots the dust temperature versus far-IR luminosity for the ALESS sample.  The solid black line shows the far-IR-luminosity-to-dust-temperature relation (L-T$_d$) as found by \cite{symeonidis} for the sources with $z<2$.  For the luminosity region where our samples overlap, the mean redshift of \cite{symeonidis} sample is $z\sim$1.1, while the ALESS sample is dominated by sources at redshifts ranging from 1.5 to 3, allowing us to test if the L-T$_d$ relation evolves further with redshift, as is seen for sources at redshifts $z<0.1$.  By choosing a fixed dust emissivity of 1.5, we are able to compare our two samples.  However, we must consider the selection effects of our sample.  In the top panel of Fig.~\ref{results}, there are red and purple, dot and dashed lines denoting representative ALMA 3.5$\sigma$ detection limits for redshifts of 1, 3, 5, and 7.  In the region where our two samples overlap ($\sim$1--4$\times10^{12}$\,L$_\odot$), it is clear that this detection limit biases our sample to colder temperatures.

To test whether or not our sample is indeed cooler, we devise a method of applying the ALESS selection effects to the \cite{symeonidis} sample.  We obtained the catalogue of sources used to create the estimate of the $L$--$T_\mathrm{d}$ relation of \cite{symeonidis}, the solid black line in Fig.~\ref{results}, including source far-IR luminosities and dust temperatures.  We randomly draw $N$ objects from this source list, where $N$ is the number of sources in the list, with replacement.  We randomly assign to these sources, redshifts from the ALESS source catalogue, such that they will have the same redshift distribution.  We retain those sources that have a predicted flux density greater than the 3.5\,$\sigma$ ALMA flux limit at 870\,$\mu$m and calculate the mean dust temperature of this sample of sources.  We perform this procedure many times, thus bootstrapping the sample, and restrict our test to sources with luminosities between $10^{12}$ and $10^{13}\,\mathrm{L}_\odot$ (where the two samples overlap).  We find a mean dust temperature of (35.6$\pm$0.8)\,K.  Using a similar bootstrapping procedure, we find a mean dust temperature of (33.7$\pm$4.2)\,K for the ALESS sample.  Since these values are consistent, we cannot conclude that we detect any evolution in dust temperature with redshift in the ALESS sources when compared to those of \cite{symeonidis}, despite the apparent difference in Fig.~\ref{results}.  The selection effects of the ALESS sample unfortunately preclude any attempt at performing this same test for those sources with $z<0.1$.  For the full sample, we find an average dust temperature of (33.9$\pm$2.4)\,K.

\subsection{Contribution to the comoving star formation rate density of the Universe}

\begin{figure*}
\begin{center}
\includegraphics[width=10cm,trim = 25mm 0mm 15mm 10mm, clip, angle=-90]{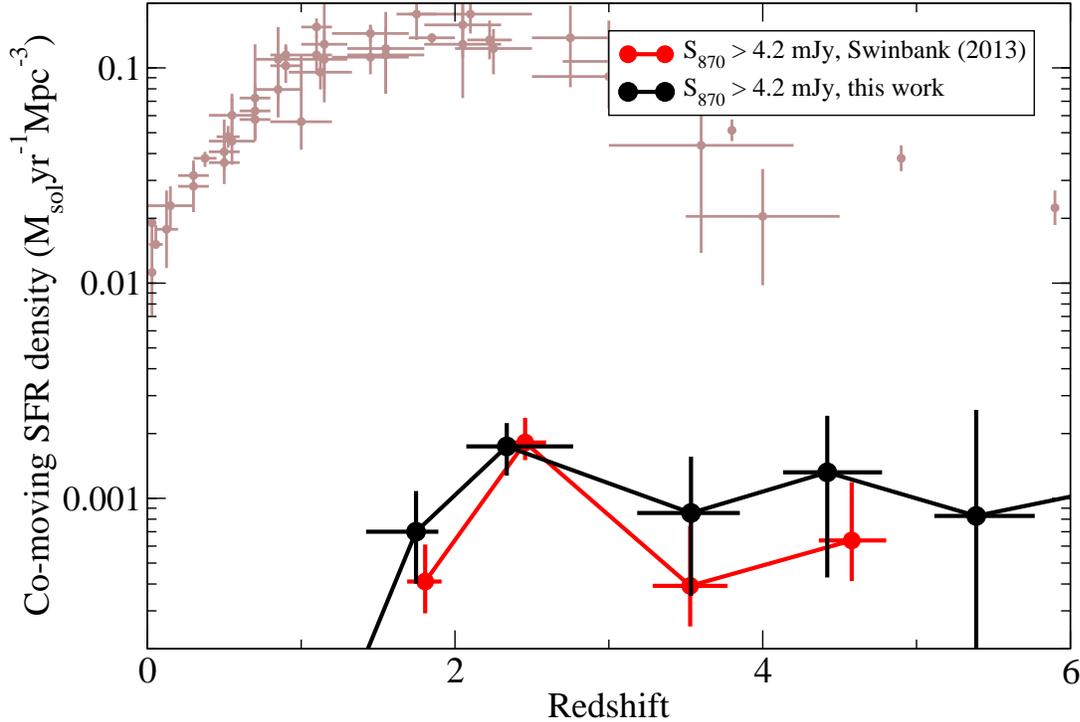}
\end{center}
\caption[Contribution of the ALESS sources with flux densities greater than 4.2\,mJy to the comoving star-formation history of the Universe.]{Contribution of the ALESS sources with flux densities greater than 4.2\,mJy to the comoving star-formation history of the Universe.  The vertical error bars on our results are 68 percent confidence intervals for the comoving SFR density after bootstrapping the MCMC chains, while the horizontal error bars are the 16th and 84th percentiles of the redshift distribution used to generate each data point, with the data point being plotted at the 50th percentile of the redshift distribution used within that bin.  For comparison, we include the estimates from \cite{swinbank}, divided by a factor of 2, for this region being underdense \citep{casey2009}.  Also included is the comoving star-formation history of the Universe as compiled by \cite{madau}.}
\label{sfrhistory2}
\end{figure*}

Fig.~\ref{sfrhistory2} shows the comoving star-formation rate (SFR) density for the ALESS sources with flux densities greater than 4.2\,mJy, using a conversion factor of $1.08\times 10^{-10} \mathrm{M}_\odot \mathrm{yr}^{-1} \mathrm{L}_\odot^{-1}$ for a Chabrier
IMF, as in \cite{swinbank}.  The vertical error bars on our results are 68 percent confidence intervals for the comoving SFR density after bootstrapping the MCMC chains and the horizontal error bars are the 16th and 84th percentile of the redshift distribution used to generate each data point, with the data point being plotted at the 50th percentile of the redshift distribution used within that bin.  For comparison, the points plotted from \cite{swinbank} are divided by a factor of 2, which they use to correct their estimate to compensate for the fact that the region is though to be underdense \citep{casey2009}.  Although our competing methods may produce significantly different far-IR constraints for individual sources, our new estimate agrees rather well with those of \cite{swinbank}.  As such, we refer the reader to \cite{swinbank} for interpretations of what this means for the role that submm galaxies play in overall star-formation at high redshift.

\section{Conclusions}

After generalising our method from M14 for the case of deblending SEDs of confused point sources, we have been able to show that our method gives realistic estimates of far-IR properties and their uncertainties, and accurately captures the degeneracies among SED parameters of nearby sources caused by confusion.  When applied to the ALESS catalogue, we were able to derive constraints on dust temperatures and far-IR luminosities and show that our results correlate with those of \cite{swinbank}, although our derived far-IR properties differ significantly when comparing individual sources.  \herschel{}-SPIRE currently provides the best view of the 250, 350 and 500\,$\mu$m extragalactic sky in terms of depth and sky coverage.  Using the sample of \cite{symeonidis} and applying the same selection function as for the ALESS sample, we show that any apparent evolution of the $L$--$T_\mathrm{d}$ relation to cooler dust temperatures at high redshifts, may be driven by selection effects.

With the large quantities of confusion limited imaging now available, such as that from \herschel{}, applications of our method are many.  One possibility is obvious:  the comoving SFR density of the Universe as seen within well studied regions, such as the Hubble Ultra Deep Field (HUDF).  Confusion limited \herschel{}-SPIRE imaging for this field are already available, and in Section~\ref{degeneracies} we showed how ALMA observations at more than one frequency can greatly aid in deblending SEDs.  Combining these observations with the spectroscopic and photometric catalogues that currently exist would yield valuable constraints on the rest-frame far-IR properties that efficiently and systematically uses all the available information in the submm images.

\section*{Acknowledgements}

This research has been supported by the Natural Sciences and Engineering Research Council of Canada.
The ALMA observations were carried out under program 2011.0.00294.S. ALMA is a partnership of ESO (representing its member states), NSF (USA) and NINS (Japan), together with NRC (Canada) and NSC and ASIAA (Taiwan), in cooperation with the Republic of Chile. The Joint ALMA Observatory is operated by ESO, AUI/NRAO and NAOJ.
SPIRE has been developed by a consortium of institutes led by Cardiff University (UK) and including Univ. Lethbridge (Canada); NAOC (China); CEA, LAM (France); IFSI, Univ. Padua (Italy); IAC (Spain); Stockholm Observatory (Sweden); Imperial College London, RAL, UCL-MSSL, UKATC, Univ. Sussex (UK); and Caltech, JPL, NHSC, Univ. Colorado (USA). This development has been supported by national funding agencies: CSA (Canada); NAOC (China); CEA, CNES, CNRS (France); ASI (Italy); MCINN (Spain); SNSB (Sweden); STFC, UKSA (UK); and NASA (USA).
PACS has been developed by a consortium of institutes led by MPE (Germany) and including UVIE (Austria); KU Leuven, CSL, IMEC (Belgium); CEA, LAM (France); MPIA (Germany); INAF-IFSI/OAA/OAP/OAT, LENS, SISSA (Italy); IAC (Spain). This development has been supported by the funding agencies BMVIT (Austria), ESA-PRODEX (Belgium), CEA/ CNES (France), DLR (Germany), ASI/INAF (Italy), and CICYT/MCYT (Spain).
This research has made use of data from the HerMES project (\url{http://hermes.sussex.ac.uk/}), a Herschel Key Programme utilising Guaranteed Time from the SPIRE instrument team, ESAC scientists and a mission scientist. 
The HerMES data were accessed through the HeDaM database (\url{http://hedam.oamp.fr}) operated by CeSAM and hosted by the Laboratoire d'Astrophysique de Marseille.
This research was enabled in part by support provided by WestGrid (\url{www.westgrid.ca}) and Compute Canada Calcul Canada (\url{www.computecanada.ca}).
This work is based [in part] on observations made with the Spitzer Space Telescope, which is operated by the Jet Propulsion Laboratory, California Institute of Technology under a contract with NASA. Support for this work was provided by NASA.
The VLA is part of the US National Radio Astronomy Observatory, a facility of the National Science Foundation operated under cooperative agreement by Associated Universities, Inc.

\bibliography{references}

\onecolumn
\begin{center}
\begin{longtable}{ccccc} 
\caption[The model fit parameters and credible intervals for the ALESS sample.]{The model fit parameters and credible intervals for the ALESS sample.  The ALMA flux density estimates are those of \cite{alma} and the photometeric redshift estimates are those of \cite{simpson}.  We report the median values from our MCMC chains for the far-IR luminosities and dust temperatures.  We report 68 percent credible intervals for both dust temperatures and far-IR luminosities.  In the case where the lower credible interval is either zero for the far-IR luminosity or 10\,K for the dust temperature, we report the 84 percent upper credible interval as an upper limit.}
\label{propertytable}\\
\hline
Gal ID  & ALMA 870\,$\mu$m & $z_{\mathrm{phot}}$ & Far-IR & Dust temp.\\ 
 & (mJy) & & luminosity (L$_\odot$) & (K)\\
\hline\\
\endfirsthead
\endhead
\hline\\
Gal ID  & ALMA 870\,$\mu$m & $z_{\mathrm{phot}}$ & Far-IR & Dust temp.\\ 
 & (mJy) & & luminosity (L$_\odot$) & (K)\\
\hline\\
\endhead
ALESS001.1 & $6.75\pm0.49$ & $4.34\substack{+2.66\\ -1.43}$	 & $9.0\substack{+5.6\\ -8.1}\times10^{12}$  & $37\substack{+12\\ -14}$ \\
ALESS001.2 & $3.48\pm0.43$ & $4.65\substack{+2.34\\ -1.02}$	 & $8.1\substack{+4.0\\ -6.8}\times10^{12}$  & $46\substack{+14\\ -14}$ \\
ALESS001.3 & $1.89\pm0.42$ & $2.85\substack{+0.20\\ -0.30}$	 & $1.2\substack{+0.5\\ -0.6}\times10^{12}$  & $29\substack{+4\\ -3}$ \\
ALESS002.1 & $3.81\pm0.42$ & $1.96\substack{+0.27\\ -0.20}$	 & $2.1\substack{+0.8\\ -2.0}\times10^{12}$  & $30\substack{+7\\ -11}$ \\
ALESS002.2 & $4.23\pm0.67$ & $3.92\substack{+0.48\\ -1.42}$	 & $2.1\substack{+2.3\\ -2.0}\times10^{12}$  & $<39$ \\
ALESS003.1 & $8.28\pm0.40$ & $3.90\substack{+0.50\\ -0.59}$	 & $1.1\substack{+0.3\\ -0.4}\times10^{13}$  & $38\substack{+5\\ -4}$ \\
ALESS005.1 & $7.78\pm0.68$ & $2.86\substack{+0.05\\ -0.04}$	 & $5.3\substack{+0.6\\ -0.6}\times10^{12}$  & $30\substack{+1\\ -1}$ \\
ALESS006.1 & $5.98\pm0.41$ & $0.45\substack{+0.06\\ -0.04}$	 & $4.3\substack{+1.1\\ -1.6}\times10^{10}$  & $11\substack{+1\\ -1}$ \\
ALESS007.1 & $6.10\pm0.32$ & $2.50\substack{+0.12\\ -0.16}$	 & $8.8\substack{+1.1\\ -1.1}\times10^{12}$  & $34\substack{+1\\ -1}$ \\
ALESS009.1 & $8.75\pm0.47$ & $4.50\substack{+0.54\\ -2.33}$	 & $<1.8\times10^{13}$  & $36\substack{+13\\ -10}$ \\
ALESS010.1 & $5.25\pm0.50$ & $2.02\substack{+0.09\\ -0.09}$	 & $3.6\substack{+0.2\\ -0.2}\times10^{12}$  & $31\substack{+1\\ -1}$ \\
ALESS011.1 & $7.29\pm0.41$ & $2.83\substack{+1.88\\ -0.50}$	 & $1.6\substack{+0.8\\ -1.3}\times10^{13}$  & $43\substack{+8\\ -15}$ \\
ALESS013.1 & $8.01\pm0.59$ & $3.25\substack{+0.64\\ -0.46}$	 & $5.2\substack{+1.6\\ -2.1}\times10^{12}$  & $30\substack{+3\\ -4}$ \\
ALESS014.1 & $7.47\pm0.52$ & $4.47\substack{+2.54\\ -0.88}$	 & $3.5\substack{+1.6\\ -2.4}\times10^{13}$  & $54\substack{+11\\ -15}$ \\
ALESS015.1 & $9.01\pm0.37$ & $1.93\substack{+0.62\\ -0.33}$	 & $3.1\substack{+1.3\\ -1.9}\times10^{12}$  & $25\substack{+4\\ -4}$ \\
ALESS015.3 & $1.95\pm0.52$ & $3.15\substack{+0.65\\ -0.65}$	 & $7.8\substack{+3.6\\ -5.6}\times10^{11}$  & $26\substack{+8\\ -7}$ \\
ALESS017.1 & $8.44\pm0.46$ & $1.51\substack{+0.10\\ -0.07}$	 & $2.2\substack{+0.2\\ -0.3}\times10^{12}$  & $24\substack{+1\\ -1}$ \\
ALESS018.1 & $4.38\pm0.54$ & $2.04\substack{+0.10\\ -0.06}$	 & $4.3\substack{+1.2\\ -1.0}\times10^{12}$  & $35\substack{+3\\ -2}$ \\
ALESS019.1 & $4.98\pm0.42$ & $2.41\substack{+0.17\\ -0.11}$	 & $3.7\substack{+0.5\\ -0.5}\times10^{12}$  & $32\substack{+1\\ -1}$ \\
ALESS019.2 & $1.98\pm0.47$ & $2.17\substack{+0.09\\ -0.10}$	 & $1.4\substack{+0.3\\ -0.3}\times10^{12}$  & $29\substack{+2\\ -2}$ \\
ALESS022.1 & $4.48\pm0.54$ & $1.88\substack{+0.18\\ -0.23}$	 & $3.4\substack{+0.8\\ -0.9}\times10^{12}$  & $30\substack{+2\\ -2}$ \\
ALESS023.1 & $6.74\pm0.37$ & $4.99\substack{+2.01\\ -2.55}$	 & $<2.7\times10^{13}$  & $50\substack{+17\\ -19}$ \\
ALESS023.7 & $1.76\pm0.49$ & $2.90\substack{+1.20\\ -0.40}$	 & $1.4\substack{+0.8\\ -1.3}\times10^{12}$  & $33\substack{+13\\ -11}$ \\
ALESS025.1 & $6.21\pm0.47$ & $2.24\substack{+0.07\\ -0.17}$	 & $5.4\substack{+0.7\\ -0.6}\times10^{12}$  & $33\substack{+1\\ -1}$ \\
ALESS029.1 & $5.90\pm0.43$ & $2.66\substack{+2.94\\ -0.76}$	 & $<2.2\times10^{13}$  & $44\substack{+14\\ -20}$ \\
ALESS031.1 & $8.12\pm0.37$ & $2.89\substack{+1.80\\ -0.41}$	 & $1.1\substack{+0.6\\ -0.8}\times10^{13}$  & $40\substack{+8\\ -12}$ \\
ALESS037.1 & $2.92\pm0.41$ & $3.53\substack{+0.56\\ -0.31}$	 & $6.7\substack{+1.9\\ -2.5}\times10^{12}$  & $44\substack{+5\\ -5}$ \\
ALESS037.2 & $1.65\pm0.44$ & $4.87\substack{+0.21\\ -0.40}$	 & $1.2\substack{+0.4\\ -0.4}\times10^{13}$  & $64\substack{+6\\ -6}$ \\
ALESS039.1 & $4.33\pm0.34$ & $2.44\substack{+0.17\\ -0.23}$	 & $2.9\substack{+0.6\\ -0.6}\times10^{12}$  & $30\substack{+2\\ -2}$ \\
ALESS041.1 & $4.88\pm0.61$ & $2.75\substack{+4.25\\ -0.72}$	 & $<4.6\times10^{13}$  & $62\substack{+18\\ -28}$ \\
ALESS041.3 & $2.68\pm0.75$ & $3.10\substack{+1.30\\ -0.60}$	 & $1.5\substack{+0.7\\ -1.1}\times10^{12}$  & $28\substack{+9\\ -8}$ \\
ALESS043.1 & $2.30\pm0.42$ & $1.71\substack{+0.20\\ -0.12}$	 & $1.0\substack{+0.2\\ -0.3}\times10^{12}$  & $28\substack{+2\\ -2}$ \\
ALESS045.1 & $6.03\pm0.54$ & $2.34\substack{+0.26\\ -0.67}$	 & $3.0\substack{+1.5\\ -1.5}\times10^{12}$  & $28\substack{+4\\ -4}$ \\
ALESS049.1 & $6.00\pm0.68$ & $2.76\substack{+0.11\\ -0.14}$	 & $7.2\substack{+0.9\\ -1.0}\times10^{12}$  & $37\substack{+2\\ -2}$ \\
ALESS049.2 & $1.80\pm0.46$ & $1.47\substack{+0.07\\ -0.10}$	 & $1.3\substack{+0.2\\ -0.3}\times10^{12}$  & $31\substack{+2\\ -2}$ \\
ALESS051.1 & $4.70\pm0.39$ & $1.22\substack{+0.03\\ -0.06}$	 & $5.5\substack{+0.8\\ -0.8}\times10^{11}$  & $20\substack{+1\\ -1}$ \\
ALESS055.1 & $3.99\pm0.36$ & $2.05\substack{+0.15\\ -0.13}$	 & $3.1\substack{+1.6\\ -1.5}\times10^{11}$  & $<18$ \\
ALESS055.2 & $2.35\pm0.60$ & $4.20\substack{+0.50\\ -0.90}$	 & $7.3\substack{+3.0\\ -4.2}\times10^{11}$  & $<21$ \\
ALESS055.5 & $1.37\pm0.37$ & $2.35\substack{+0.11\\ -0.13}$	 & $4.4\substack{+1.7\\ -3.9}\times10^{11}$  & $26\substack{+7\\ -7}$ \\
ALESS057.1 & $3.56\pm0.61$ & $2.95\substack{+0.05\\ -0.10}$	 & $5.9\substack{+0.6\\ -0.7}\times10^{12}$  & $40\substack{+2\\ -2}$ \\
ALESS059.2 & $1.94\pm0.44$ & $2.09\substack{+0.78\\ -0.29}$	 & $1.2\substack{+0.6\\ -0.8}\times10^{12}$  & $31\substack{+6\\ -6}$ \\
ALESS061.1 & $4.29\pm0.51$ & $6.52\substack{+0.36\\ -0.34}$	 & $2.2\substack{+0.3\\ -0.3}\times10^{13}$  & $60\substack{+3\\ -3}$ \\
ALESS063.1 & $5.59\pm0.35$ & $1.87\substack{+0.10\\ -0.33}$	 & $1.1\substack{+0.3\\ -0.3}\times10^{12}$  & $22\substack{+2\\ -2}$ \\
ALESS065.1 & $4.16\pm0.43$ & $2.82\substack{+0.95\\ -0.36}$	 & $5.0\substack{+1.8\\ -2.6}\times10^{12}$  & $35\substack{+5\\ -6}$ \\
ALESS066.1 & $2.50\pm0.48$ & $2.33\substack{+0.05\\ -0.04}$	 & $6.0\substack{+0.4\\ -0.4}\times10^{12}$  & $42\substack{+1\\ -1}$ \\
ALESS067.1 & $4.50\pm0.38$ & $2.14\substack{+0.05\\ -0.09}$	 & $1.1\substack{+0.6\\ -0.9}\times10^{12}$  & $23\substack{+4\\ -13}$ \\
ALESS067.2 & $1.73\pm0.41$ & $2.05\substack{+0.06\\ -0.16}$	 & $3.3\substack{+1.8\\ -2.7}\times10^{11}$  & $22\substack{+7\\ -7}$ \\
ALESS068.1 & $3.70\pm0.56$ & $3.60\substack{+1.10\\ -1.10}$	 & $5.8\substack{+1.0\\ -1.0}\times10^{12}$  & $42\substack{+2\\ -2}$ \\
ALESS069.1 & $4.85\pm0.63$ & $2.34\substack{+0.27\\ -0.44}$	 & $2.3\substack{+0.8\\ -0.8}\times10^{12}$  & $29\substack{+3\\ -3}$ \\
ALESS069.2 & $2.36\pm0.56$ & $4.75\substack{+0.35\\ -1.05}$	 & $9.4\substack{+3.4\\ -5.2}\times10^{11}$  & $<19$ \\
ALESS069.3 & $2.05\pm0.56$ & $4.80\substack{+0.30\\ -1.10}$	 & $8.7\substack{+3.7\\ -5.5}\times10^{11}$  & $<23$ \\
ALESS070.1 & $5.23\pm0.45$ & $2.28\substack{+0.05\\ -0.06}$	 & $7.6\substack{+0.5\\ -0.5}\times10^{12}$  & $36\substack{+1\\ -1}$ \\
ALESS071.1 & $2.85\pm0.60$ & $2.48\substack{+0.21\\ -0.11}$	 & $1.7\substack{+0.2\\ -0.3}\times10^{13}$  & $49\substack{+2\\ -2}$ \\
ALESS071.3 & $1.36\pm0.38$ & $2.73\substack{+0.22\\ -0.25}$	 & $1.1\substack{+0.5\\ -0.5}\times10^{12}$  & $35\substack{+4\\ -5}$ \\
ALESS072.1 & $4.91\pm0.50$ & $4.15\substack{+0.55\\ -1.65}$	 & $5.4\substack{+2.5\\ -4.9}\times10^{12}$  & $37\substack{+10\\ -8}$ \\
ALESS073.1 & $6.09\pm0.47$ & $5.18\substack{+0.43\\ -0.45}$	 & $7.6\substack{+1.6\\ -1.7}\times10^{12}$  & $38\substack{+4\\ -3}$ \\
ALESS074.1 & $4.64\pm0.69$ & $1.80\substack{+0.13\\ -0.13}$	 & $2.4\substack{+0.2\\ -0.2}\times10^{12}$  & $30\substack{+1\\ -1}$ \\
ALESS075.1 & $3.17\pm0.45$ & $2.39\substack{+0.08\\ -0.06}$	 & $5.8\substack{+0.4\\ -0.5}\times10^{12}$  & $36\substack{+1\\ -1}$ \\
ALESS075.4 & $1.30\pm0.37$ & $2.10\substack{+0.29\\ -0.34}$	 & $5.7\substack{+1.9\\ -2.4}\times10^{11}$  & $23\substack{+3\\ -3}$ \\
ALESS076.1 & $6.42\pm0.58$ & $4.50\substack{+0.20\\ -2.00}$	 & $<6.1\times10^{12}$  & $33\substack{+10\\ -7}$ \\
ALESS079.1 & $4.12\pm0.37$ & $2.04\substack{+0.63\\ -0.31}$	 & $2.1\substack{+1.0\\ -1.3}\times10^{12}$  & $29\substack{+5\\ -5}$ \\
ALESS079.2 & $1.98\pm0.40$ & $1.55\substack{+0.11\\ -0.18}$	 & $1.5\substack{+0.6\\ -0.6}\times10^{12}$  & $33\substack{+4\\ -4}$ \\
ALESS079.4 & $1.81\pm0.51$ & $4.60\substack{+1.20\\ -0.60}$	 & $1.2\substack{+0.6\\ -0.9}\times10^{12}$  & $<31$ \\
ALESS080.1 & $4.03\pm0.86$ & $1.96\substack{+0.16\\ -0.14}$	 & $1.1\substack{+0.3\\ -0.4}\times10^{12}$  & $23\substack{+2\\ -2}$ \\
ALESS080.2 & $3.54\pm0.90$ & $1.37\substack{+0.17\\ -0.08}$	 & $4.6\substack{+1.6\\ -1.8}\times10^{11}$  & $19\substack{+2\\ -2}$ \\
ALESS082.1 & $1.93\pm0.47$ & $2.10\substack{+3.27\\ -0.44}$	 & $8.0\substack{+6.1\\ -7.7}\times10^{12}$  & $56\substack{+16\\ -26}$ \\
ALESS084.1 & $3.17\pm0.63$ & $1.92\substack{+0.09\\ -0.07}$	 & $1.6\substack{+0.7\\ -1.4}\times10^{12}$  & $28\substack{+9\\ -6}$ \\
ALESS084.2 & $3.25\pm0.77$ & $1.75\substack{+0.08\\ -0.19}$	 & $1.0\substack{+0.3\\ -0.3}\times10^{12}$  & $26\substack{+3\\ -3}$ \\
ALESS087.1 & $1.34\pm0.35$ & $3.20\substack{+0.08\\ -0.47}$	 & $1.0\substack{+0.2\\ -0.2}\times10^{13}$  & $58\substack{+5\\ -5}$ \\
ALESS087.3 & $2.44\pm0.59$ & $4.00\substack{+1.10\\ -0.30}$	 & $2.5\substack{+0.9\\ -1.3}\times10^{12}$  & $33\substack{+6\\ -6}$ \\
ALESS088.1 & $4.62\pm0.58$ & $1.84\substack{+0.12\\ -0.11}$	 & $1.1\substack{+0.5\\ -0.5}\times10^{12}$  & $22\substack{+4\\ -3}$ \\
ALESS088.2 & $2.14\pm0.50$ & $5.20\substack{+0.60\\ -1.20}$	 & $1.5\substack{+0.7\\ -1.0}\times10^{12}$  & $<32$ \\
ALESS088.5 & $2.86\pm0.72$ & $2.30\substack{+0.11\\ -0.50}$	 & $3.7\substack{+1.2\\ -1.1}\times10^{12}$  & $37\substack{+5\\ -3}$ \\
ALESS088.11 & $2.51\pm0.71$ & $2.57\substack{+0.04\\ -0.12}$	 & $7.2\substack{+4.0\\ -6.0}\times10^{11}$  & $24\substack{+8\\ -8}$ \\
ALESS092.2 & $2.42\pm0.68$ & $1.90\substack{+0.28\\ -0.75}$	 & $1.3\substack{+0.6\\ -1.0}\times10^{11}$  & $<17$ \\
ALESS094.1 & $3.18\pm0.52$ & $2.87\substack{+0.37\\ -0.64}$	 & $3.5\substack{+1.3\\ -1.5}\times10^{12}$  & $35\substack{+5\\ -4}$ \\
ALESS098.1 & $4.78\pm0.60$ & $1.63\substack{+0.17\\ -0.09}$	 & $7.2\substack{+1.1\\ -1.5}\times10^{12}$  & $33\substack{+1\\ -2}$ \\
ALESS099.1 & $2.05\pm0.43$ & $5.00\substack{+1.20\\ -0.60}$	 & $1.5\substack{+0.6\\ -0.9}\times10^{12}$  & $<25$ \\
ALESS102.1 & $3.08\pm0.50$ & $1.76\substack{+0.16\\ -0.18}$	 & $1.3\substack{+0.3\\ -0.3}\times10^{12}$  & $26\substack{+2\\ -1}$ \\
ALESS103.3 & $1.43\pm0.41$ & $4.40\substack{+0.70\\ -0.70}$	 & $1.5\substack{+1.0\\ -1.3}\times10^{12}$  & $38\substack{+11\\ -22}$ \\
ALESS107.1 & $1.91\pm0.39$ & $3.75\substack{+0.09\\ -0.08}$	 & $4.9\substack{+2.1\\ -1.8}\times10^{12}$  & $47\substack{+7\\ -5}$ \\
ALESS107.3 & $1.46\pm0.40$ & $2.12\substack{+1.54\\ -0.81}$	 & $<17.6\times10^{11}$  & $31\substack{+11\\ -16}$ \\
ALESS110.1 & $4.11\pm0.47$ & $2.55\substack{+0.70\\ -0.50}$	 & $5.5\substack{+2.2\\ -3.2}\times10^{12}$  & $41\substack{+6\\ -7}$ \\
ALESS110.5 & $2.39\pm0.60$ & $3.70\substack{+0.40\\ -1.20}$	 & $4.4\substack{+1.6\\ -2.6}\times10^{11}$  & $<16$ \\
ALESS112.1 & $7.62\pm0.49$ & $1.95\substack{+0.15\\ -0.26}$	 & $2.8\substack{+0.7\\ -0.7}\times10^{12}$  & $27\substack{+2\\ -2}$ \\
ALESS114.1 & $2.99\pm0.78$ & $3.00\substack{+1.40\\ -0.50}$	 & $1.1\substack{+0.5\\ -0.8}\times10^{13}$  & $46\substack{+8\\ -12}$ \\
ALESS114.2 & $1.98\pm0.50$ & $1.56\substack{+0.07\\ -0.07}$	 & $4.2\substack{+0.3\\ -0.3}\times10^{12}$  & $36\substack{+1\\ -1}$ \\
ALESS116.1 & $3.08\pm0.47$ & $3.54\substack{+1.47\\ -0.87}$	 & $3.3\substack{+2.0\\ -3.0}\times10^{12}$  & $36\substack{+13\\ -14}$ \\
ALESS116.2 & $3.42\pm0.57$ & $4.02\substack{+1.19\\ -2.19}$	 & $<8.5\times10^{12}$  & $40\substack{+16\\ -14}$ \\
ALESS118.1 & $3.20\pm0.54$ & $2.26\substack{+0.50\\ -0.23}$	 & $2.5\substack{+0.9\\ -1.2}\times10^{12}$  & $33\substack{+5\\ -5}$ \\
ALESS119.1 & $8.27\pm0.54$ & $3.50\substack{+0.95\\ -0.35}$	 & $1.1\substack{+0.3\\ -0.5}\times10^{13}$  & $39\substack{+5\\ -6}$ \\
ALESS122.1 & $3.69\pm0.42$ & $2.06\substack{+0.05\\ -0.06}$	 & $8.5\substack{+0.6\\ -0.6}\times10^{12}$  & $38\substack{+1\\ -1}$ \\
ALESS124.1 & $3.64\pm0.57$ & $6.07\substack{+0.94\\ -1.16}$	 & $5.3\substack{+3.0\\ -4.0}\times10^{12}$  & $<47$ \\
ALESS124.4 & $2.24\pm0.58$ & $5.60\substack{+0.60\\ -1.20}$	 & $5.2\substack{+2.2\\ -3.1}\times10^{12}$  & $45\substack{+9\\ -9}$ \\
ALESS126.1 & $2.23\pm0.55$ & $1.82\substack{+0.28\\ -0.08}$	 & $8.4\substack{+1.9\\ -2.3}\times10^{11}$  & $30\substack{+3\\ -3}$ \\

\hline
\end{longtable}
\end{center}
\twocolumn
%\end{document}

\end{document}